\documentclass[sigconf,nonacm]{acmart}
\AtBeginDocument{%
  }

\usepackage[linesnumbered,ruled,vlined]{algorithm2e}
\usepackage{enumitem}
\usepackage{bbding}
\usepackage{diagbox}
\usepackage{graphicx}
\usepackage{pifont}
\usepackage{hyperref}
\newcommand{\logtext}[1]{``\textls[-30]{\texttt{#1}}''}

\newcommand{\method}{LogBatcher}
\newcommand{\tool}{LogBatcher }
\newcommand{\metricone}{$\mathbf{T}_{\text{total}}$}
\newcommand{\metrictwo}{$\mathbf{T}_{\text{invoc}}$}

\usepackage{xcolor}
\definecolor{mygreen}{HTML}{548235}
\definecolor{myorange}{HTML}{C55A11}

\begin{document}

\title{Stronger, Cheaper and Demonstration-Free Log Parsing with LLMs}

\author{Yi Xiao}
\affiliation{
    \institution{Chongqing University}
    \city{Chongqing}
    \country{China}}
\email{yixiao@cqu.edu.cn}

\author{Van-Hoang Le}
\affiliation{
    \institution{The University of Newcastle}
    \city{NSW}
    \country{Australia}}
\email{vanhoang.le@uon.edu.au}

\author{Hongyu Zhang}
\authornote{Hongyu Zhang is the corresponding author}
\affiliation{
    \institution{Chongqing University}
    \city{Chongqing}
    \country{China}}
\email{hyzhang@cqu.edu.cn}


\begin{abstract}

Log parsing, the process of converting raw log messages into structured formats, is an important initial step for automated analysis of logs of large-scale software systems. Traditional log parsers often rely on heuristics or handcrafted features, which may not generalize well across diverse log sources or require extensive model tuning. 
Recently, some log parsers have utilized powerful generative capabilities of large language models (LLMs). However, they heavily rely on demonstration examples, resulting in substantial overhead in LLM invocations. To address these issues, we propose \method, a cost-effective LLM-based log parser that requires no training process or labeled data. To leverage latent characteristics of log data and reduce the overhead, we divide logs into several partitions through clustering. Then we perform a cache matching process to match logs with previously parsed log templates. Finally, we provide LLMs with better prompt context specialized for log parsing by batching a group of logs from each partition. We have conducted experiments on 16 public log datasets and the results show that \tool is effective and efficient for log parsing.

\end{abstract}


\maketitle

\section{Introduction}
\label{sec:introduction}
Software-intensive systems often record runtime information by printing console logs. Software logs are semi-structured data printed by logging statements (e.g., \texttt{printf()}, \texttt{logInfo()}) in source code. The primary purpose of system logs is to record system states and important events at various critical points to help engineers better understand system behaviours and diagnose problems. The rich information included in log data enables a variety of software reliability management tasks, such as 
detecting system anomalies~\cite{zhang2019robust, le2022log}, ensuring application security~\cite{milajerdi2019poirot, oprea2015detection}, and diagnosing errors~\cite{jia2017logsed, le2021log}. 

\begin{figure}[h]
    \centering
    \includegraphics[width=.8\linewidth]{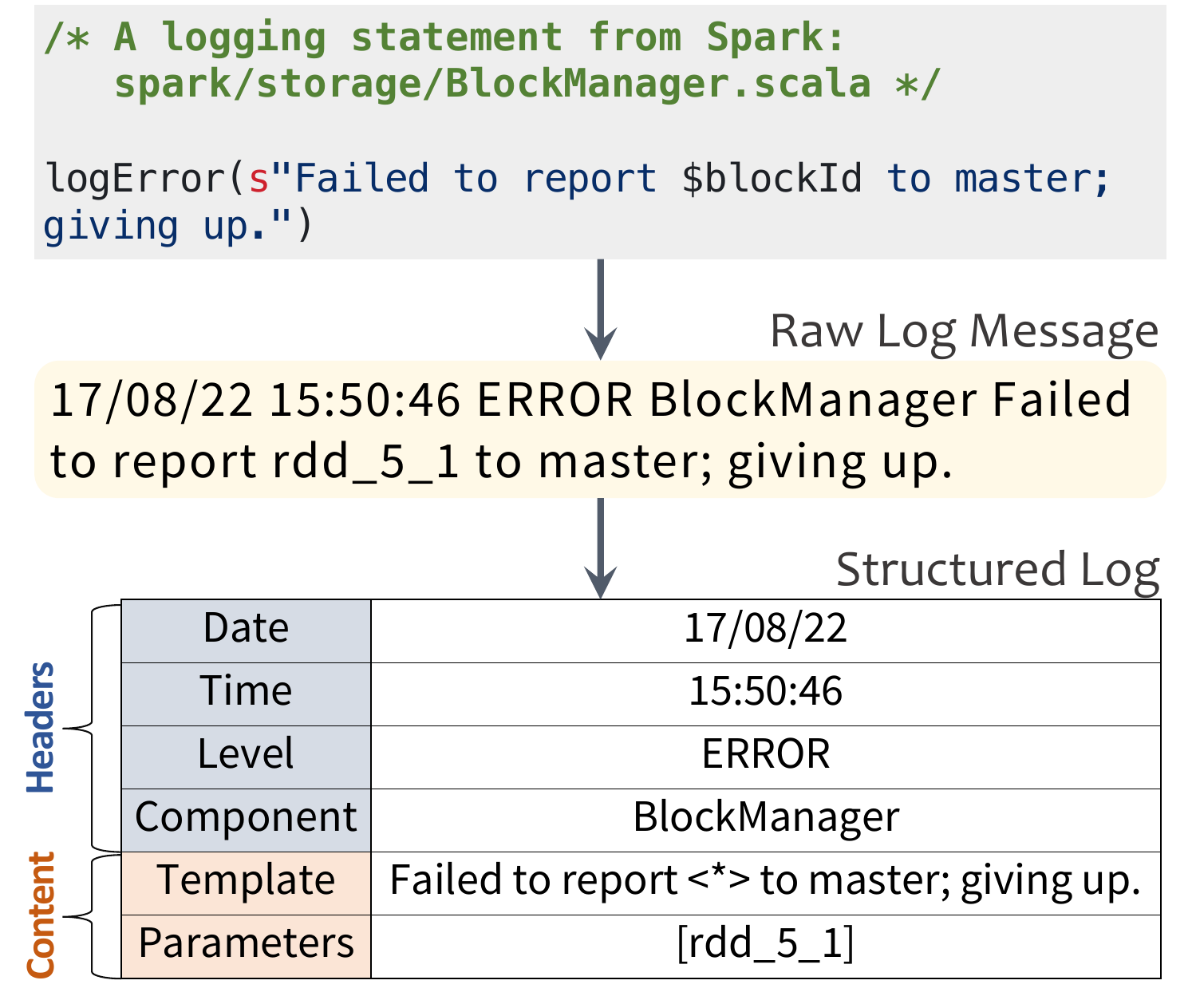}
    \caption{An Illustration of Log Parsing}
    \label{fig:log-parsing-example}
\end{figure}

To facilitate various downstream analytics tasks, log parsing, which parses free-text into a structured format~\cite{zhu2019tools}, is the first and foremost step. An accurate log parser is always in high demand for intelligent log analytics because it could simplify the process of downstream analytics tasks and allow more methods (e.g., Machine Learning and Deep Learning) to be applied~\cite{he2020survey}.
Log parsing is the task of converting a raw log message into a specific log template associated with the corresponding parameters. 
As shown in Figure~\ref{fig:log-parsing-example}, each log message is printed by a logging statement in the source code and records a specific system event with its header and body. The header is determined by the logging framework and includes information such as component and verbosity level. The log message body (log message for short) typically consists of two parts: 1) \textit{Template} - constant strings (or keywords) describing the system event; 2) \textit{Parameters} - dynamic variables, which vary during runtime and reflect system runtime information. For example, in the log message in Figure~\ref{fig:log-parsing-example}, the header (i.e., \logtext{17/08/22 15:50:46}, \logtext{INFO}, and \logtext{BlockManager}) can be easily distinguished through regular expressions. The log message consists of a template \logtext{Failed to report <*> to master; giving up} and a parameter \logtext{rdd\_5\_1}. The log template typically contains constant strings, referring to commonalities across log data. The log parameters are dynamic variables, referring to variabilities that vary across log messages.

In recent years, there have been tremendous efforts towards achieving the goal of automated log parsing. Since the source code is generally inaccessible during system maintenance, existing log parsing methods propose to leverage \textit{syntax} and \textit{semantic} patterns of logs to identify and separate static text and dynamic variables. Syntax-based log parsers~\cite{he2017drain, du2016spell, dai2020logram, makanju2009clustering} utilize specific features or heuristics (e.g., token count, frequency, and position) to extract the constant parts of log messages as templates. In contrast, semantic-based log parsers propose to recognize dynamic variables based on their semantic differences from constant keywords. Unfortunately, the performance of these log parsers in practice remains unsatisfactory~\cite{jiang2024large, petrescu2023log}. On the one hand, syntax-based log parsers heavily rely on crafted rules and domain knowledge, thus being ineffective when encountering previously unseen log patterns~\cite{le2023log, jiang2023lilac}. On the other hand, semantic-based log parsers still require certain training overheads, such as training models from scratch or fine-tuning pre-trained language models with labeled data, which is scarce and costly to obtain~\cite{le2023evaluation}.

To address these limitations, recent studies~\cite{le2023evaluation, jiang2023lilac, xu2024divlog} propose to leverage the text understanding capacity of large language models (LLMs) for automated log parsing. Specifically, these studies adopt the in-context learning (ICL) prompting technique to adapt LLMs to the log parsing task. In ICL, a prompt consists of an instruction and associated demonstration examples. 
Despite the effectiveness, these LLM-based log parsers still fail to meet practical usage of log parsing due to the following reasons:

\begin{enumerate}
    \item \textbf{Over reliance on demonstrations:} As LLMs are not explicitly specialized for log parsing, existing LLM-based log parsers require labeled demonstration examples (i.e., demonstrations) to construct in-context prompts. 
    The performance of LLM-based log parsing has been shown to be sensitive to the quality and quantity of demonstrations~\cite{jiang2023lilac, le2023evaluation}. Furthermore, demonstrations can be quickly outdated as the volume and format of logs rapidly change~\cite{kabinna2018examining, zhang2019robust}. 
    Hence, selecting demonstrations in in-context learning can be a delicate art and might require significant trial-and-errors. 
    \item \textbf{LLM invocation cost:} 
   Log data is typically generated in a massive volume. Naively querying LLMs for each log message is impractical due to the substantial cost of invoking LLMs' service API. Furthermore, the cost incurred by the instruction and demonstrations in the prompts is not neglectable.
 
\end{enumerate}

To address the aforementioned challenges, in this paper, we propose LogBatcher, a novel \textit{training-free}, \textit{demonstration-free}, and \textit{cost-effective} LLM-based log parser. \tool leverages latent commonalities and variabilities of log data~\cite{li2024logshrink} to provide LLMs with better prompt context specialized for log parsing.
Specifically, \tool first groups log data into several partitions using a versatile clustering algorithm. 
Then, for each partition, \tool samples log messages with high diversity to construct a batch of logs as the prompt to query LLMs to parse logs. By doing so, we can introduce variabilities within the prompt context to better guide LLMs to perform the log parsing task without the need for demonstrations. To further reduce the number of LLM invocations, \tool adopts a simple yet effective caching mechanism to store the intermediate results of LLMs and avoid redundant queries.

We have conducted a comprehensive evaluation on the public LogPai dataset~\cite{zhu2019tools}. The results show that \tool outperforms state-of-the-art baselines in terms of both accuracy and LLM inference cost. It can achieve an average Group Accuracy~\cite{zhu2019tools} of 0.972 and Message-Level Accuracy of 0.895, which are significantly higher than the best-performing supervised LLM-based log parser (i.e., LILAC~\cite{jiang2023lilac}). Moreover, \tool is robust across diverse log datasets without the need for demonstrations, and can substantially reduce the cost of LLM invocation by at least 106\%.


The main contributions of this paper are as follows:
\begin{enumerate}
    \item We propose LogBatcher, the first \textit{demonstration-free} LLM-based log parsing framework to the best of our knowledge. Besides, \tool does not require any training overhead and is cost-effective for parsing large-scale log data.
    \item We introduce a log-specific prompting strategy to provide LLMs with a batch of logs, which allows LLMs to better incorporate the latent commonalities and variabilities among log messages. Furthermore, the token consumption of LLMs is reduced. 
    \item We conduct a comprehensive evaluation on the public LogPai dataset~\cite{zhu2019tools}. Experimental results show that \tool outperforms state-of-the-art baselines in terms of both accuracy and LLM invocation cost.
\end{enumerate}

The remainder of this paper is organized as follows. Section~\ref{sec:background} provides background and related works on log parsing and LLMs. Section~\ref{sec:motivation} presents a motivating example to illustrate the challenges of existing LLM-based log parsing methods. Section~\ref{sec:methodology} described our proposed \tool framework in detail. Section~\ref{sec:experimental-design} states the experimental design for evaluating LogBatcher. Section~\ref{sec:sec:experimental-results} presents the experimental results. Section~\ref{sec:discussion} discusses the results and implications, followed by Section~\ref{sec:conclusion} to conclude the paper.

\section{Background and Related Work}
\label{sec:background}
\subsection{Log Parsing}
Log parsing is one of the first steps for log analysis tasks~\cite{zhu2019tools}. It is a process to extract the static log template parts and the corresponding dynamic parameters (or variables) from free-text raw log messages. A straightforward method of log parsing involves matching raw log messages with logging statements within the source code~\cite{xu2009detecting, nagappan2009efficiently} or designing handcrafted regular expressions to extract log templates and parameters~\cite{zhu2019tools}.
However, these approaches is impractical due to the inaccessibility of the source code (especially for third-party libraries~\cite{zhu2019tools}) and the huge volume of logs. 
To achieve the goal of automated log parsing, many \textit{syntax}-based and \textit{semantic}-based approaches have been proposed to identify log templates as the frequent part of log messages. 
Syntax-based log parsers~\cite{he2017drain, yu2023brain, dai2020logram, jiang2008abstracting} assume that log templates inherit some common patterns which emerge constantly across the entire log dataset.
Some parsers~\cite{dai2020logram, vaarandi2015logcluster, nagappan2010abstracting} extract log templates by identifying the constant parts of log messages through the mining of frequent patterns, for example, Logram~\cite{dai2020logram} finds frequent $n$-gram patterns which emerge constantly across the entire log dataset as templates. 
Logs that belong to the same template exhibit similarities. Consequently, some methods~\cite{fu2009execution,  tang2011logsig, shima2016length} employ clustering techniques to group logs and extract the constant portions of log messages for log parsing. 
Heuristics-based log parsers~\cite{jiang2008abstracting, he2017drain, yu2023brain} leverage unique characteristics from log messages to extract common templates efficiently. For example, AEL~\cite{jiang2008abstracting} employs a list of heuristic rules to extract common templates. Drain~\cite{he2017drain} employs a fixed-depth tree structure to assist in dividing logs into different groups, assuming that all log parameters within specific templates possess an identical number of tokens, while Brain~\cite{yu2023brain} updated Drain by using a bidirectional parallel tree.




Semantic-based log parsers leverage semantic differences between keywords and parameters to formulate log parsing as a token classification task. For example, UniParser~\cite{liu2022uniparser} unifies log parsing for heterogeneous log data by training with labeled data from multiple log sources to capture common patterns of templates and parameters. 
LogPPT~\cite{le2023log} introduces a novel paradigm for log parsing, employing template-free prompt-tuning to fine-tune the pre-trained language model, RoBERTa. 
Although effective, existing semantic-based log parsers require certain training overheads, such as training models from scratch or fine-tuning pre-trained language models with labeled data, which is scarce and costly to obtain~\cite{le2023evaluation}.

Recently, some studies have proposed to utilize large language models (LLMs) owing to their extensive pre-trained knowledge. These studies have achieved promising results in log parsing~\cite{le2023evaluation, jiang2023lilac, xu2024divlog} thanks to the strong in-context learning capability of LLMs. In the following sections, we will introduce some recent LLM-based log parsers and discuss their limitations.




\subsection{Log Parsing with Large Language Models}
\label{sec:llm-log-parsing}

Large language models (LLMs) have achieved remarkable success in various natural language processing~\cite{devlin2018bert, lewis2019bart} and computer vision tasks~\cite{guo2023images, shao2023prompting}. 
In-context learning is a promising prompt engineering method for adopting LLMs without fine-tuning them~\cite{liu2023pre}. 
In-context learning typically requires an \textit{instruction} that describes the task and \textit{demonstrations} that provide several examples of how to solve the task. Recent studies have demonstrated that in-context learning can aid LLMs in achieving remarkable performance in a variety of tasks~\cite{wei2022chain, khot2022decomposed, xia2024fuzz4all}. 

Le and Zhang~\cite{le2023evaluation} validated the potential of LLMs in log parsing and obtained promising results. DivLog~\cite{xu2024divlog} and LILAC~\cite{jiang2023lilac} enhance the performance of large models by selecting demonstrations from labeled log data and utilizing the in-context learning capabilities of LLMs. They employ different methods to sample a labeled candidate log set. 
These methods are sensitive to the quantity and coverage of labeled logs and incur LLM inference overhead. Lemur~\cite{zhang2024lemur} invokes LLM to merge generated similar templates, improving the accuracy of log parsing groupings. However, it requires extensive hyperparameter tuning for specific datasets.
It has been found that these LLM-based log parsers have outperformed semantic-based log parsers (e.g., LogPPT~\cite{le2023log} and UniParser~\cite{liu2022uniparser}) in terms of parsing accuracy~\cite{jiang2023lilac, xu2024divlog}.

Despite promising results, LLM-based log parsing can be costly in terms of token usage, especially when large volumes of LLM calls are needed. The costs of one LLM invocation scale linearly with the number of tokens, including both the input prompt tokens (instruction and demonstrations). 
Consequently, managing LLM invocation cost is vital for practical applications. Since LLM infrastructure/services can change over time, recent studies~\cite{cheng2023batch, hidvegi2024cigar} measure and reduce token consumption as the primary metric for LLM cost management. Similarly, in this paper, we focus on accomplishing more data processing with fewer tokens and LLMs calls to achieve cost-effective log parsing.


\section{A Motivating Example}
\label{sec:motivation}
Recently, several studies~\cite{le2023evaluation, jiang2023lilac, xu2024divlog} have proposed to utilize LLMs for log parsing and achieved promising results. Still, these studies fail to achieve satisfactory performance in practice. We have identified two major limitations of existing LLM-based log parsing approaches, which prevent their practical usage.

\begin{figure}[ht]
    \centering
    \includegraphics[width=\linewidth]{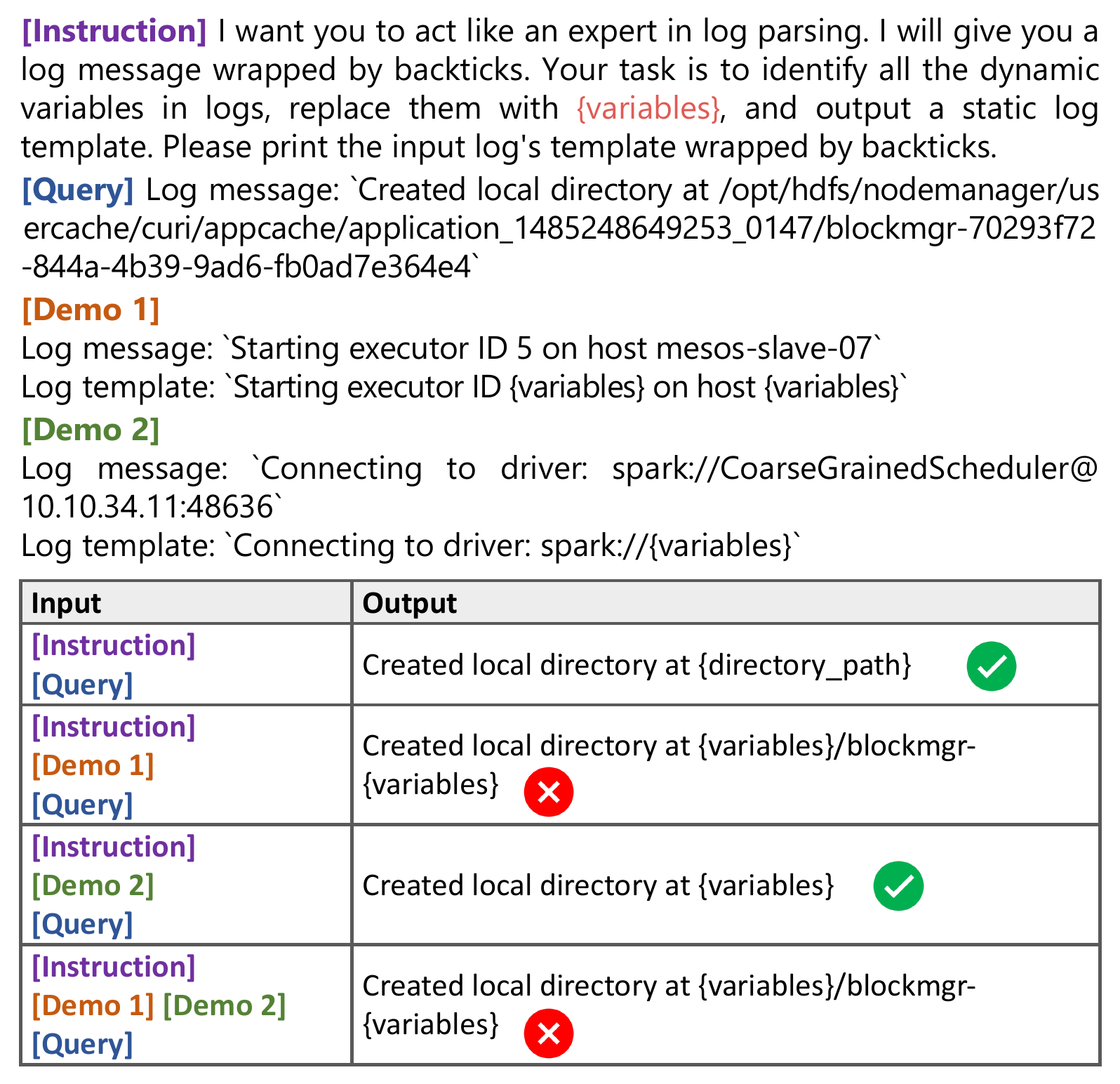}
    \vspace{-6pt}
    \caption{Selecting in-context demonstrations for log parsing on Spark (\textit{Results are produced using gpt-3.5-turbo~\cite{gpt-3.5-turbo} with instruction and demonstrations adopted from~\cite{jiang2023lilac}})}
    \vspace{-6pt}
    \label{fig:log-parsing-icl-example}
\end{figure}

\textit{Over reliance on demonstrations.}
Although LLMs are equipped with a huge amount of pre-trained knowledge, they are not specialized in the log parsing task. Directly querying LLMs for log parsing could result in unsatisfactory performance~\cite{jiang2023lilac, le2023evaluation}. Hence, to overcome this problem, recent studies~\cite{xu2024divlog, jiang2023lilac} straightforwardly leverage the in-context learning prompting technique to impart log-specific knowledge to LLMs via labeled demonstrations. 
However, selecting even a few useful demonstrations can quickly become more laborious as the volume and format of logs rapidly change~\cite{kabinna2018examining, zhang2019robust}. 
More importantly, selecting in-context demonstrations can be 
challenging 
as the quality of these demonstrations directly affects LLM-based log parsing. Figure~\ref{fig:log-parsing-icl-example} illustrates the impact of four different demonstrations on the parsing performance. Sample inputs and outputs shown from top to bottom (Spark log) are: (1) zero-shot without demonstration: correct answer; (2) a correct but noisy demonstration (\textcolor{myorange}{Demo 1}), which leads to a wrong answer; (3) a correct demonstration (\textcolor{mygreen}{Demo 2}), which leads to a correct answer; and (4) combining \textcolor{myorange}{Demo 1} and \textcolor{mygreen}{Demo 2} again leads to an incorrect answer. 
This issue highlights the sensitivity of demonstrations to the performance of LLM-based log parsing.


\textit{LLM invocation cost.}
Log data can be generated in a massive volume in production. For example, Mi et al.~\cite{mi2013toward} reported that the Alibaba cloud system produces about 30-50 gigabytes (around 100-200 million lines) of tracing logs per hour. 
Naively querying LLMs for each log message is impractical due to the substantial cost of inference. 
As illustrated in Figure~\ref{fig:log-parsing-icl-example}, querying GPT-3.5-Turbo~\cite{gpt-3.5-turbo} with 
a prompt consisting of one instruction (66 tokens in total\footnote{We compute the number of tokens using the OpenAI's \textit{tiktoken} package: \url{https://github.com/openai/tiktoken}}) and one log message (55 tokens in total) will cost (55 $+$ 66) $\times$ 100,000,000 $\times$ (0.50$/$1,000,000) = \$6,050 because the price of GPT-3.5-Turbo API services is \$0.5 per 1M tokens\footnote{\url{https://openai.com/pricing}}. 
Due to the large amount of log messages, the cost for LLM-based log parsing could pose a significant financial burden in practice. 

Note that the cost incurred by the instruction and demonstrations in the prompt is not neglectable. 
For example, the token count of the prompt instruction in Figure~\ref{fig:log-parsing-icl-example} is 66, which is more than the token count of the log message (55). Considering adding more demonstrations (36 tokens per demonstration in Figure~\ref{fig:log-parsing-icl-example}) to the prompt to improve the parsing performance, the token count of the prompt will increase linearly with the number of demonstrations. This will further increase the cost of LLM invocation, making it even more expensive to query LLMs for log parsing.




\section{Methodology}
\label{sec:methodology}

\begin{figure*}[t]
    \centering
    \includegraphics[width=\textwidth]{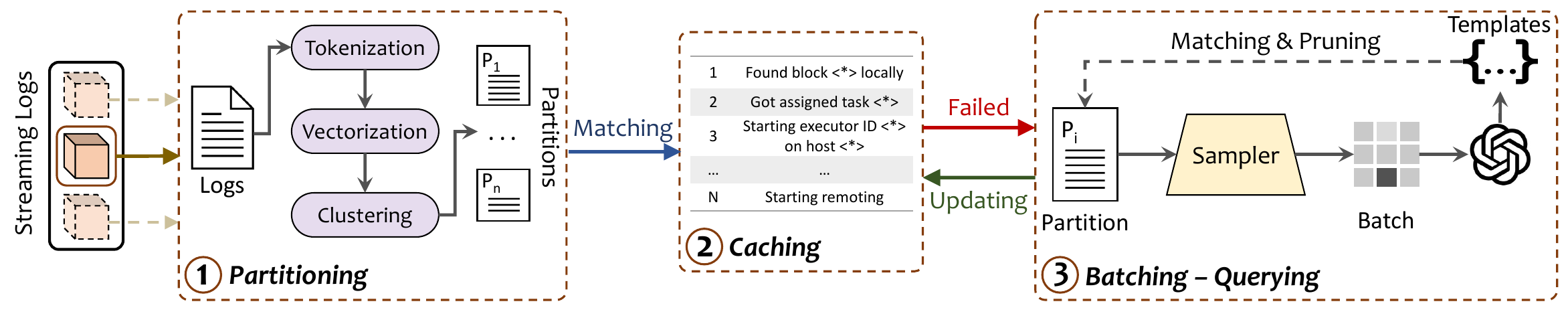} 
    \caption{An overview of \tool}
    \label{fig:overflow}
\end{figure*}

Drawing upon the observations described in Section~\ref{sec:motivation}, we propose LogBatcher, a novel \textit{demonstration-free},  \textit{training-free}, and \textit{cost-effective} LLM-based log parser. The main idea behind \tool is that log data possesses latent characteristics, i.e., commonality and variability, 
which allow LLMs to perform log parsing without demonstrations.
Specifically, as the goal of log parsing is to recognize the dynamic variables (i.e., variability) from static patterns (i.e., commonality), we use a batch of log messages as the input to LLMs instead of using a single log message. In this way, we can incorporate commonalities and variabilities among log messages into the input of LLM, thus allowing LLMs to better correlate the log parsing with the log data itself without the need of labeled demonstration examples.

An overview of \tool framework is illustrated in Figure~\ref{fig:overflow}. Since raw log data are massively generated in the production environment~\cite{mi2013toward, wang2022spine}, we divide the raw log data into multiple chunks before analysis. Log chunks are processed in parallel, each log chunk goes through three main components: \ding{172} \textit{Partitioning}: separating each log chunk into several partitions using a versatile clustering algorithm. \ding{173} \textit{Caching}: performing a cache matching process for logs in each partition to match them with previously parsed log templates to avoid duplicate LLM queries and improve parsing efficiency. \ding{174} \textit{Batching -- Querying}: sampling a diverse set of logs from each partition to form a batch, which is then sent to the LLM for parsing. 
Finally, we refine the identified templates and match the logs with the templates to mitigate the impact of clustering errors. 



\subsection{Partitioning}
\label{subsec:partitioning}

\begin{figure}[h]
    \centering
    \vspace{-6pt}
    \includegraphics[width=.95\linewidth]{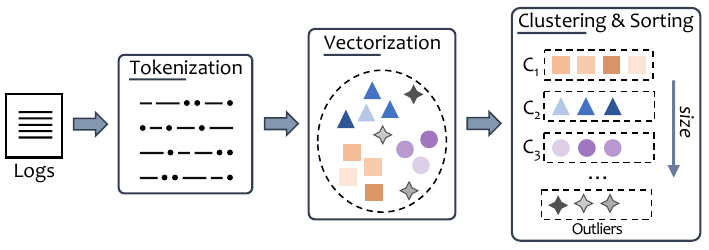}
    \vspace{-6pt}
    \caption{Log partitioning through clustering}
    \vspace{-6pt}
    \label{fig:clustering}
\end{figure}

The aim of this phase is to ensure that logs allocated to the same partitions share some commonalities. This is crucial for the subsequent in-context learning process, as it allows LLMs to learn the commonalities within log data and associate them with the log parsing task. We employ a versatile clustering algorithm based on DBSCAN~\cite{ester1996density} to partition logs. Figure~\ref{fig:clustering} illustrates the log partitioning process.

\subsubsection{Tokenization}

The initial step of \textit{partitioning} involves log tokenization and cleaning, which are crucial for accurate clustering. 
First, we use general delimiters (i.e., \textit{white space}) to perform initial tokenization of the logs. Considering that logs have some unique delimiters due to their relevance to the code, we define specific rules to further refine the tokenization for each token. Finally, we clean the logs by masking potential variable tokens.
We utilize some basic regular expressions to refine the tokenization. For example, the symbol \logtext{=} can serve as a delimiter in logs such as \logtext{START: tftp pid=16563 from=10.100.4.251}. However, if \logtext{=} appears within a URL, as in \logtext{after trim url = https://www.google.com/search?q=test}, it disrupts the integrity of the variable, leading to clustering errors. After that, we improve the clustering performance by masking tokens that resemble parameters such as numbers, IP addresses, and URLs. Given a batch of logs $\mathcal {\text{L}} = \{L_1, L_2, \ldots, L_n\}$, each log $L_i$ is tokenized into a set of tokens $\{t_{i1}, t_{i2}, \ldots, t_{im}\}$.



\subsubsection{Vectorization}
Vectorization is a prerequisite for clustering as it transforms log data into a numerical format, which is suitable for clustering algorithms. Since different tokens in logs are of varying importance~\cite{zhang2019robust}, we adopt the Frequency-Inverse Document Frequency (TF-IDF)~\cite{salton1988term} to vectorize the log. Specifically, we first calculate the \textit{Term Frequency} (TF) to describe the importance of a token in a log message, where $TF(token) = \frac{\#token}{\#total}$, $\#token$ is the number of target token in a log message, $\#total$ is the number of all tokens in a log message. On the other hand, if a token appears in many logs, it is less informative and becomes too common to be able to distinguish distinct log messages. Therefore, we calculate the \textit{Inverse Document Frequency} (IDF) to reduce the weight of overly common tokens, where $IDF(token) = \log(\frac{\#L}{\#L_{token}})$, $\#L$ is the total number of logs, $\#L_{token}$ is the number of logs containing the target token. For each word, its TF-IDF weight $w$ is calculated by $TF \times IDF$.

Finally, we can obtain the vector representation $\mathbf{V}_L \in \mathbb{R}^d$ of each log message by summing up the token vectors $L$ with their corresponding TF-IDF weights, according to Equation~\ref{eq:TF-IDF}:

\begin{equation}\label{eq:TF-IDF}
    \mathbf{V}_L = \frac{1}{N} \sum_{i=1}^{N} w_i~.~\mathit{v}_i
\end{equation}


\subsubsection{Clustering \& Sorting}
\tool adopts the DBSCAN algorithm (Density-Based Spatial Clustering of Applications with Noise)~\cite{ester1996density} to cluster log messages in a chunk into different groups, each of which is more likely to contain the log messages with similar semantics. DBSCAN groups together data points that are closely packed, marking as outliers points that lie alone in low-density regions. The reasons we choose DBSCAN are threefold: (1) it does not require specifying the number of clusters in advance, which is more practical in the log parsing task; (2) it has been demonstrated to be more effective and efficient and has been widely used in many domains~\cite{schubert2017dbscan}; (3) it has a small number of hyperparameters 
and is less sensitive to hyperparameter selection, thus being easy-to-use in practice.
After clustering, we sort the clusters by size in descending order and consider all outliers as a separate cluster to process at last. 
The reason is that smaller clusters are more likely to contain logs with unique characteristics (e.g., noises) that are difficult to be parsed. By processing them at last, we can leverage previously parsed templates stored in the cache to filter out the noise and improve parsing performance.

\subsection{Caching} \label{subsec: caching}
Parsing all arriving logs with LLMs is impractical due to the high API cost and latency, especially when logs are generated in large quantities. To address this issue and improve parsing efficiency, we leverage a simple caching mechanism to store previously parsed log templates and match them with logs in the current partition. Specifically, before parsing, we filter out logs that can be matched with the cache. For those unmatched logs, we process them with LLMs, adding the newly generated template into the cache. Each new item in the cache contains three values: (1) a newly generated template from LLM, (2) a reference log that can match the template, and (3) the matching frequency (how many logs the template has matched). We detail the usage of these values below.


To match logs with the template, some log parsers~\cite{he2017drain, jiang2023lilac} select logs with a similarity above a certain threshold to the template and consider them as matches, which could result in mismatches. 
In our approach, we perform regular expression matching. Specifically, this involves replacing \logtext{<*>} in the log templates with the generic matching symbol \logtext{(.?)}, allowing regular expressions to check if the logs and templates match exactly and return all the corresponding variables. Additionally, inspired by~\cite{shima2016length}, the reference log is used to verify whether its length is consistent with that of the target log, making our caching more precise. To enhance caching efficiency, we also dynamically sort the templates in the cache so that the frequently occurring templates can be checked first.



\subsection{Batching -- Querying}
Logs that belong to the same template not only share frequently occurring tokens but also exhibit rich variability in their dynamic parts. 
These characteristics of log data are widely observed in practice, and are adopted by many data-driven log parsing methods~\cite{he2017drain, shima2016length}. Recent LLM-based log parsers, however, overlook these characteristics, leading to the overly sensitive nature of LLMs to demonstrations. To address this issue, we propose a batching - querying approach to provide LLMs with commonalities and variabilities within input logs for demonstration-free log parsing.
\subsubsection{Batching}
After partitioning, logs in each partition already exhibit commonalities in their semantics and syntax. 
We sample a 
set of logs from each partition to form an input batch for LLMs.
To this end, we adopt a diversity-based sampling method to select logs that maximize the sample diversity. Specifically, we calculate the cosine similarity between every two logs based on their TF-IDF vectors, forming a similarity matrix. We then use the Determinantal Point Process (DPP) algorithm~\cite{kulesza2012dpp} 
to select logs that maximize the sample diversity. By doing so, we can ensure that the input batch contains both commonalities (introducing by clustering-based partitioning) and variabilities (introducing by diversity-based sampling) within the input logs, which can help LLMs better associate the task description with the input logs and improve parsing accuracy.

\subsubsection{Prompting Design}
\begin{figure}[h]
    \centering
    \includegraphics[width=\linewidth]{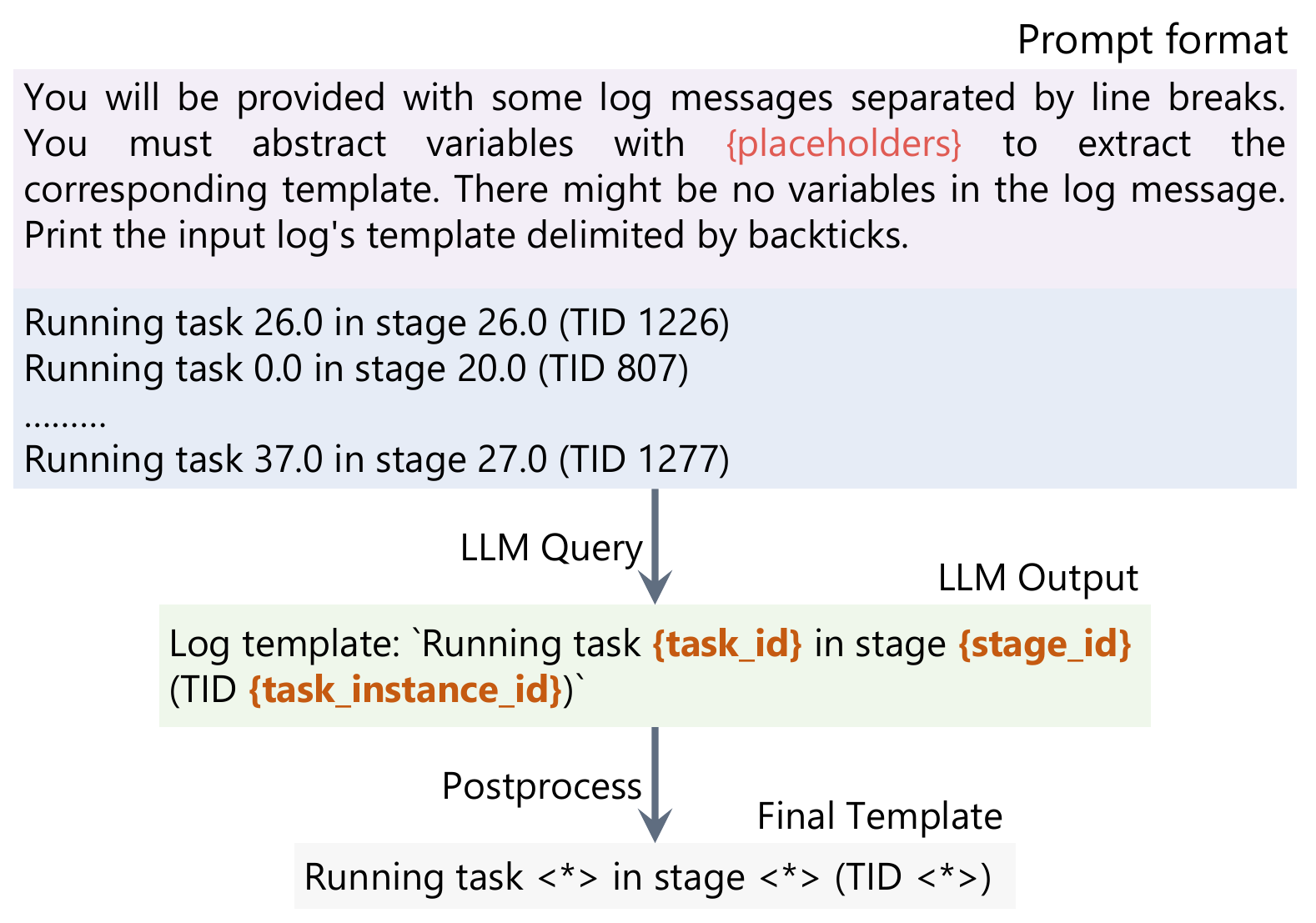}
    \caption{An illustration of our prompt design}
    \label{fig:prompt-format}
\end{figure}

A common in-context learning paradigm consists of three parts: instruction, demonstration and query.
Since our method is demonstration-free, our prompt consists only of instruction and query.
Following previous work~\cite{le2023evaluation, jiang2023lilac}, we design and use the prompt format, as shown in Figure~\ref{fig:prompt-format}. However, different from them, we provide LLMs with the input in the form of a batch as follows:
\begin{enumerate}[leftmargin=*]
    \item \textit{Instruction}: To provide the LLM with task-specific information, we briefly describe the goal of log parsing, and the formats of input and output. Moreover, we emphasize the main objective of log parsing as abstracting the variables as well as indicate that logs may not contain variables to avoid over-parsing, where the LLM tries to find variables in every log.
    \item \textit{Queried Log Batch}: We provide the LLM with a batch of logs as input, separated by the newline character. This batch is sampled from the partitioned logs, which contain both commonalities and variabilities within the input logs. Hence, the input is well-related to the instruction, which can help LLMs better understand the log parsing task.
\end{enumerate}


\subsubsection{Post-Processing}
The output from an LLM may contain redundant information beyond the desired template. With the locator \logtext{\`} and placeholder \logtext{\{placeholder\}}, we can easily filter the raw output from LLM and get the identified template. To make the style of labeling the same for every system, Khan et al.~\cite{khan2022guidelines} customized some heuristic rules, to correct the identified template. Some related works~\cite{jiang2023lilac, le2023log} also adopt these rules to refine the generated templates and minimize the impact of inconsistent labels. We adopt and optimize this post-process. For example, \cite{khan2022guidelines} only considers decimal numbers in the logs as variables, but in reality, hexadecimal numbers appear just as frequently. 

\subsubsection{Matching \& Pruning}
For the results of clustering, two common issues usually arise: logs that belong to the same template are grouped into different clusters, and logs that belong to different clusters are mistakenly grouped into the same cluster. The methods mentioned earlier effectively solve the first problem. For the second problem, we use the matching \& pruning method. Pruning is essentially a re-group process using the identified template. As mentioned in Section~\ref{subsec: caching}, the identified template can be matched with logs through transformation with regular expressions. In most cases, the resulting template can match all logs in the cluster. When not all logs can be matched, indicating the second issue mentioned earlier in clustering, we consider the template as valid for the logs it can match. The unmatched logs are then pruned and sent to a new cluster, which will reenter the queue sequence for further parsing.
Even though logs may be misclassified into the same cluster and trigger an invocation, we still make good use of this invocation, avoiding additional overhead. Algorithm~\ref{alg:matching-pruning} shows this process.

\begin{algorithm}
    \small
    \SetAlgoLined
    \DontPrintSemicolon
    \caption{Matching \& Pruning}
    \label{alg:matching-pruning}
    \KwIn{$\mathcal{C}$: Parsed Cluster \\
    \hspace*{3.1em} $T$: Identified Template}
    \KwOut{$\mathcal{C}_{new}$: New Cluster}
    \BlankLine
    $\text{regex} \gets \text{convertToRegex}(T)$\;
    $\mathcal{C}_{new} \gets \emptyset$\;
    \ForEach{$\text{log} \in \mathcal{C}$}{
        \If{$\text{match}(\text{log}, \text{regex})$}{
            $\mathcal{C} \gets \mathcal{C} \setminus \{\text{log}\}$\;
            $\mathcal{C}_{new} \gets \mathcal{C}_{new} \cup \{\text{log}\}$\;
        }
    }
    \Return{$\mathcal{C}_{new}$}\;

\end{algorithm}

After this batching - querying process, we obtain a log template that can match all or partial (if pruning is needed) logs in each partition. We repeat this process for every partition that cannot find corresponding template, until all partitions are successfully parsed.




\section{Experimental Design}
\label{sec:experimental-design}
\subsection{Research Questions}
We evaluate our approach by answering the following research questions:

\textbf{RQ1. How does \tool perform compared to the baselines?}
In this RQ, we aim to comprehensively evaluate the performance of our proposed method.
Specially, we compare our method with four state-of-the-art unsupervised data-driven log parsers (i.e., Drain~\cite{he2017drain}, AEL~\cite{jiang2008abstracting}, Brain~\cite{yu2023brain}, and Logram~\cite{dai2020logram}) and two LLM-based supervised log parsers (i.e., DivLog~\cite{xu2024divlog} and LILAC~\cite{jiang2023lilac}).
We adopt the implementation of these methods from their public replication packages~\cite{zhu2019tools, xu2024divlog, jiang2023lilac}.
DivLog and LILAC are recently proposed to leverage the in-context learning capacity of LLMs for log parsing.
For a fair comparison, we use the same settings from LILAC~\cite{jiang2023lilac} to reproduce the results of both DivLog and LILAC, in which 32 candidates are sampled from the log data and 3 demonstrations are selected as the parsing context for each queried log.

\textbf{RQ2. How do different modules contribute to LogBatcher?}
\tool consists of three main components: Partitioning, Caching, and Batching. We evaluate the importance of each component by removing each of them from the framework and evaluating the performance. Specifically, we perform the ablation study with the following settings: (1) w/o\textsubscript{partitioning}: we divide logs into several partitions using time windows; (2) w/o\textsubscript{caching}: we directly remove the caching component; and (3) w/o\textsubscript{batching}: we only use one log entry as the LLM input.

\textbf{RQ3. How does \tool perform with demonstrations?}
Although \tool is demonstration-free, it can easily be extended to use labeled demonstration examples as other supervised LLM-based approaches do. 
In this RQ, we adopt the same setting from LILAC~\cite{jiang2023lilac} to select a few demonstrations for each LLM invocation. Specifically, we first sample 32 labeled candidate logs from each dataset and then select the most similar candidate logs as demonstrations for each LLM query.
We compare \tool with LILAC (the top-performing supervised LLM-based log parser) 
to evaluate the effectiveness of our approach with different numbers of demonstration examples. 


\textbf{RQ4. How do different settings affect LogBatcher?}
To delve deeper into the effectiveness and robustness of LogBatcher, we explore how different settings of its major components affect the overall performance. Specifically, we use different sampling methods for batching, different batch sizes, and different LLMs to evaluate the performance of LogBatcher. 

\subsection{Datasets}
We conduct experiments on 16 public log datasets (Loghub-2k~\cite{loghub2021}) originated from the LogPai project~\cite{zhu2019tools}.
These datasets cover logs from distributed systems, standalone software, supercomputers, PC operating systems, mobile systems, microservices, etc. 
Zhu et al.~\cite{zhu2019tools} sampled 2,000 log entries from each system in the dataset and manually labeled them. However, it has been observed that the original labels have some errors due to inconsistent labeling styles~\cite{khan2022guidelines,loghub2021, jiang2024large}. Therefore, following existing work~\cite{le2023log, xu2024divlog, Ma_2024}, we use the version of the datasets corrected by Khan et al.~\cite{khan2022guidelines}. Furthermore, as the Proxifier dataset has many different versions, we calibrated some labels according to the guidelines proposed by~\cite{jiang2024large}. The datasets we used in our experiments are publicly available at our webpage\footnote{\url{https://anonymous.4open.science/r/LogBatcher}}. 

\subsection{Evaluation Metrics}
Following recent studies~\cite{le2023log, liu2022uniparser, le2023evaluation}, we use three main metrics for evaluation, including:

\textbf{Group Accuracy (GA):} Group Accuracy~\cite{he2017drain} is the most commonly used metric for log parsing. The GA metric is defined as the ratio of “correctly parsed" log messages over the total number of log messages, where a log message is considered “correctly parsed" if and only if it is grouped with other log messages consistent with the ground truth.

\textbf{Message Level Accuracy (MLA):} Message Level Accuracy (or Parsing Accuracy)~\cite{liu2022uniparser} is defined as the ratio of “correctly parsed" log messages over the total number of log messages, where a log message is considered to be “correctly parsed" if and only if every token of the log message is correctly identified as template or parameter.

\textbf{Edit Distance (ED):} Edit Distance assesses the performance of template extraction in terms of string comparison~\cite{nedelkoski2020self}. It calculates the minimum number of actions needed to convert one template into another. We apply normalized Edit Distance~\cite{marzal1993computation}, which computes the mean Edit Distance of all compared template pairs in the dataset (parsed templates vs ground truth templates).

In addition, to assess the efficiency of our proposed approach, we measure the token consumption. 
Specifically, we use two metrics to measure the token consumption, including (1) \metricone: the total number of tokens consumed for all invocations when parsing a dataset and (2) \metrictwo: the average number of tokens consumed per invocation.


\subsection{Implementation and Settings}
\textit{Implementation.}
In our experiments, we set the default LLM to \textit{GPT-3.5-Turbo} (version 0125\footnote{\url{https://platform.openai.com/docs/models/gpt-3-5-turbo}}), which is widely used in recent research~\cite{le2023evaluation, xu2024divlog}. We conduct our experiments on a Ubuntu 20.04 LTS server with Python 3.8. We invoke the LLM through the Python library provided by OpenAI~\cite{openai2023gpt4}.

\noindent
\textit{Settings.}
We adopt the implementation of DBSCAN provided by sklearn~\cite{pedregosa2011scikit} for the Partitioning component. We set the hyperparameters of DBSCAN as follows: \textit{epsilon} = 0.5 and \textit{min\_samples} = 5. For the Batching component, we set the batch size to 10. To avoid the randomness of the LLM, we set the temperature to 0.

\section{Results and Analysis}
\label{sec:experimental-results}
\begin{table*}[!htbp]
  \centering
  \caption{Comparison with the state-of-the-art log parsers}
  \label{tab:rq1_overall_result}
  \resizebox{\textwidth}{!}{
    \setlength{\tabcolsep}{2.5pt}
    \renewcommand{\arraystretch}{1.17}
    \begin{tabular}{c|ccc|ccc|ccc|ccc|ccc|ccc|ccc}
    \toprule
     & \multicolumn{3}{c|}{\textbf{Drain}} & \multicolumn{3}{c|}{\textbf{AEL}} & \multicolumn{3}{c|}{\textbf{Brain}} & \multicolumn{3}{c|}{\textbf{Logram}} & \multicolumn{3}{c|}{\textbf{\textit{DivLog}}} & \multicolumn{3}{c|}{\textbf{\textit{LILAC}}} & \multicolumn{3}{c}{\textbf{\tool}}\\                                                                                                   
    \cline{2-22}
    & GA & MLA & ED & GA & MLA & ED & GA & MLA & ED & GA & MLA & ED & GA & MLA & ED & GA & MLA & ED & GA & MLA & ED  \\
    \hline
    HDFS  & 0.998  & 0.999  & 0.999  & 0.998  & 0.999  & 0.999  & 0.998  & 0.959  & 0.997  & 0.930  & 0.961  & 0.993  & 0.930  & 0.996  & 0.999  & \textbf{1.000 } & \textbf{1.000 } & \textbf{1.000 } & \textbf{1.000 } & \textbf{1.000 } & \textbf{1.000 } \\
    Hadoop & 0.948  & 0.613  & 0.882  & 0.869  & 0.606  & 0.901  & 0.950  & 0.158  & 0.751  & 0.694  & 0.195  & 0.708  & 0.683  & 0.744  & 0.915  & 0.991  & \textbf{0.958 } & \textbf{0.986 } & \textbf{0.990 } & 0.886  & 0.952  \\
    Spark & 0.920  & 0.398  & 0.963  & 0.905  & 0.398  & 0.952  & 0.998  & 0.376  & 0.950  & 0.470  & 0.296  & 0.915  & 0.738  & 0.960  & 0.983  & \textbf{0.999 } & \textbf{0.983 } & \textbf{0.998 } & 0.998  & 0.972  & 0.989  \\
    Zookeeper & 0.967  & 0.799  & 0.981  & 0.965  & 0.800  & 0.981  & 0.989  & 0.779  & 0.987  & 0.956  & 0.805  & 0.970  & 0.955  & 0.979  & 0.998  & \textbf{1.000 } & 0.987  & \textbf{0.999 } & 0.995  & \textbf{0.988 } & 0.995  \\
    BGL   & 0.963  & 0.479  & 0.885  & 0.957  & 0.474  & 0.883  & \textbf{0.996 } & 0.426  & 0.891  & 0.702  & 0.282  & 0.785  & 0.736  & 0.950  & \textbf{0.990 } & 0.983  & \textbf{0.972 } & 0.989  & 0.987  & 0.941  & 0.989  \\
    HPC   & 0.887  & 0.662  & 0.872  & 0.904  & 0.680  & 0.880  & 0.945  & 0.660  & 0.973  & \textbf{0.978 } & 0.751  & 0.870  & 0.935  & 0.980  & 0.997  & 0.970  & \textbf{0.994 } & \textbf{0.999 } & 0.953  & 0.943  & 0.995  \\
    Thunderbird & 0.957  & 0.180  & 0.941  & 0.945  & 0.180  & 0.943  & 0.971  & 0.060  & 0.932  & 0.554  & 0.097  & 0.826  & 0.234  & 0.879  & 0.978  & \textbf{0.984 } & \textbf{0.913 } & \textbf{0.983 } & 0.914  & 0.854  & 0.953  \\
    Windows & 0.997  & 0.466  & 0.948  & 0.691  & 0.158  & 0.840  & 0.997  & 0.463  & \textbf{0.976 } & 0.694  & 0.141  & 0.915  & 0.710  & 0.715  & 0.903  & 0.696  & \textbf{0.685 } & 0.897  & \textbf{1.000 } & 0.609  & 0.862  \\
    Linux & 0.422  & 0.217  & 0.750  & 0.405  & 0.205  & 0.745  & 0.358  & 0.176  & 0.770  & 0.186  & 0.125  & 0.684  & 0.484  & 0.620  & 0.935  & 0.298  & 0.422  & 0.926  & \textbf{0.998 } & \textbf{0.976 } & \textbf{0.992 } \\
    Android & 0.885  & 0.750  & \textbf{0.972 } & 0.773  & 0.540  & 0.876  & 0.960  & 0.253  & 0.924  & 0.795  & 0.436  & 0.822  & 0.737  & 0.677  & 0.952  & 0.953  & 0.627  & 0.923  & \textbf{0.971 } & \textbf{0.787 } & 0.953  \\
    HealthApp & 0.901  & 0.375  & 0.749  & 0.893  & 0.368  & 0.744  & \textbf{1.000 } & 0.261  & 0.871  & 0.833  & 0.677  & 0.850  & 0.876  & 0.984  & 0.997  & 0.998  & \textbf{0.988 } & \textbf{0.998 } & 0.920  & 0.914  & 0.961  \\
    Apache & \textbf{1.000 } & 0.978  & 0.996  & \textbf{1.000 } & 0.978  & 0.996  & \textbf{1.000 } & 0.984  & 0.996  & 1.000  & 0.972  & 0.995  & 0.984  & 0.985  & 0.997  & 1.000  & \textbf{1.000 } & \textbf{1.000 } & \textbf{1.000 } & 0.978  & 0.996  \\

    Proxifier & 0.765  &0.704  &	980  &	0.826  &	0.690  &	0.972  &	0.527  &0.704  &0.945  &0.477 	&0.816 &	0.559 	 &0.531  &	0.993 	 &\textbf{0.999}  &\textbf{1.000} 	 &\textbf{0.995}  &	\textbf{0.999} 	 &\textbf{1.000}  &	0.986  &	0.997 \\
    OpenSSH & 0.789  & 0.594  & 0.919  & 0.547  & 0.729  & 0.965  & \textbf{1.000 } & 0.287  & 0.948  & 0.802  & 0.928  & 0.960  & 0.749  & 0.987  & \textbf{0.999 } & 0.753  & 0.805  & 0.983  & \textbf{1.000 } & \textbf{0.976 } & 0.989  \\
    OpenStack & 0.224  & 0.105  & 0.693  & 0.249  & 0.034  & 0.718  & 0.492  & 0.112  & 0.937  & 0.315  & 0.071  & 0.724  & 0.220  & 0.437  & 0.873  & 1.000  & 0.977  & 0.991  & \textbf{1.000 } & \textbf{0.982 } & \textbf{0.995 } \\
    Mac   & 0.814  & 0.392  & 0.896  & 0.765  & 0.284  & 0.835  & \textbf{0.949 } & 0.383  & \textbf{0.902 } & 0.759  & 0.359  & 0.843  & 0.712  & 0.549  & 0.898  & 0.805  & \textbf{0.562 } & 0.892  & 0.831  & 0.528  & 0.879  \\
    \hline
    Average & 0.840  &	0.544  &	0.902  &	0.793  &	0.508  &	0.889  &	0.883 & 	0.440  &	0.922  &	0.700  &	0.473  &	0.855  
  &	0.701  &	0.839  &	0.963  &	0.902  &	0.867  &	\textbf{0.973} 	 & \textbf{0.972}  &	\textbf{0.895}  &	0.969 \\
    
    \bottomrule
    \end{tabular}%
   }
  \label{tab:addlabel}%
\end{table*}%

\subsection{\textbf{RQ1:} How does \tool perform compared to baselines?}


This RQ evaluates the performance of \tool from three aspects: effectiveness, robustness,  and efficiency. 

\subsubsection{Effectiveness} Table~\ref{tab:rq1_overall_result} provides a comparative analysis of various log parsing methods across multiple datasets in terms of Group Accuracy (GA), Message-Level Accuracy (MLA), and Edit Distance (ED).
For each dataset, the highest accuracy of each metric is highlighted in \textbf{bold}.
Experimental results show that \tool significantly outperforms other unsupervised log parsers, including Drain~\cite{he2017drain}, AEL~\cite{jiang2008abstracting}, Brain~\cite{yu2023brain}, and Logram~\cite{dai2020logram}. Specifically, \tool exceeds the highest GA, MLA, and ED of these methods by 8.9\%, 35.1\%, and 4.7\% on average, respectively. Compared to supervised LLM-based log parsers, \tool also achieves superior performance. Specifically, \tool significantly outperforms DivLog~\cite{xu2024divlog} in terms of all three metrics. For example, it achieves better GA on all datasets, i.e., 1.6\% (Apache) to 78.0\% (OpenStack) higher than DivLog. Compared to LILAC~\cite{jiang2023lilac}, the top-performing LLM-based log parser, \tool achieves better GA and MLA on average. It also achieves comparable results in terms of ED (0.4\% lower on average). It is worth noting that without any labeled demonstrations, \tool still can achieve best average GA and MLA, and the second-best average ED. Overall, the experimental results confirm that \tool is effective for the log parsing task.

\begin{figure*}[h]
    \centering
    \includegraphics[width=.9\linewidth]{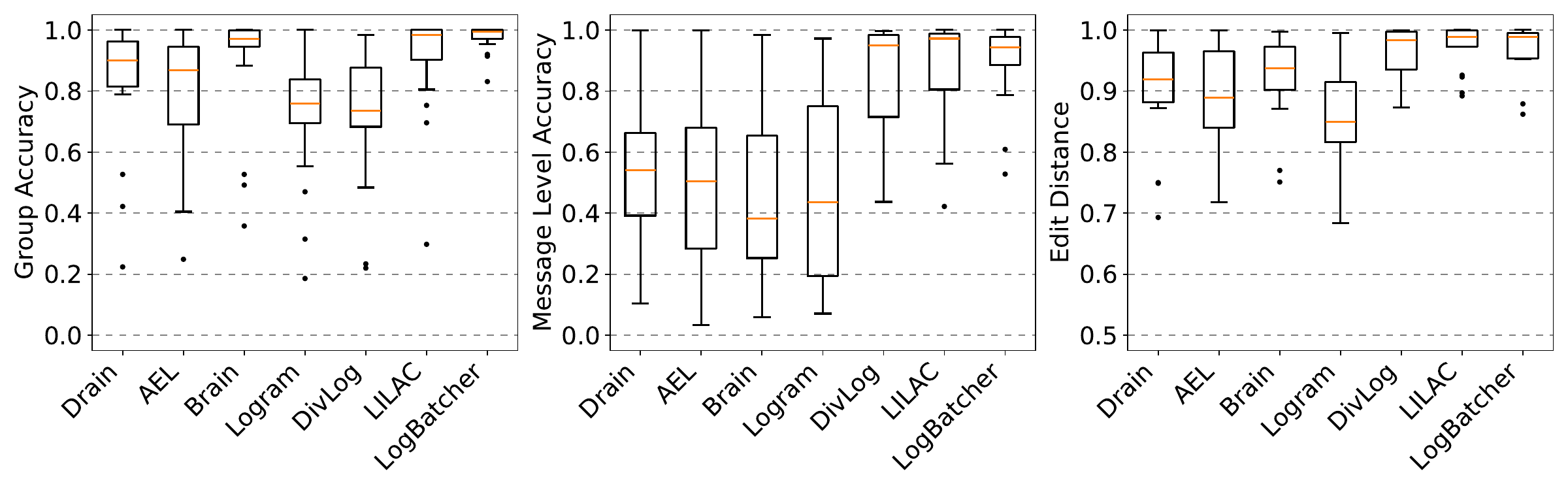 } 
    \vspace{-6pt}
    \caption{Robustness comparison between baselines and \tool}
    \vspace{-6pt}
    \label{fig:robust}
\end{figure*}

\subsubsection{Robustness} 
\tool aims to support a wide range of log data from various systems, as a universal log parser in a production environment demands strong performance and generalization capabilities~\cite{zhu2019tools}. Hence, we analyze and compare the robustness against different types of logs of \tool with that of the baselines by drawing a box plot to illustrate the accuracy distribution of each log parser's metrics across all datasets.
Figure~\ref{fig:robust} shows the results.
It is obvious that \tool consistently achieves the narrowest interquartile range (IQR) across all three metrics, indicating the stable performance of LogBatcher. Specifically, \tool yields a median of 0.99 for GA robustness, 0.94 for PA robustness and 0.99 for ED robustness, which are better or comparable to the top-performing baseline, LILAC.
Additionally, \tool's performance exhibits significantly fewer outliers compared to other baseline methods. This indicates that even in less typical scenarios, \tool can still achieve stable and reliable results.
Overall, the experimental results demonstrate that \tool is robust and can be applied to various log datasets effectively.



\begin{table}[h]
\centering
\caption{Efficiency of LLM-based Log parsers (\#tokens)}
\vspace{-6pt}
\small
\label{tab:efficiency}
\setlength{\tabcolsep}{2pt}
\renewcommand{\arraystretch}{1.1}
\begin{tabular}{c|ccc|ccc} 
\toprule
& \multicolumn{3}{c}{\metricone}  & \multicolumn{3}{c}{\metrictwo}  \\
\cline{2-7}
&  DivLog & LILAC & \tool & DivLog & LILAC & \tool \\
\hline

HDFS  & 706689  & \textbf{4793 } & 6101  & 353   & \textbf{342 } & 436  \\
Hadoop & 491299  & 33519  & \textbf{15695 } & 246   & 308   & \textbf{141 } \\
Spark & 409054  & 10587  & \textbf{4993 } & 205   & 294   & \textbf{156 } \\
Zookeeper & 360558  & 14858  & \textbf{6930 } & 180   & 310   & \textbf{122 } \\
BGL   & 420346  & 35061  & \textbf{15692 } & 210   & 297   & \textbf{139 } \\
HPC   & 288954  & 9179  & \textbf{5076 } & 144   & 255   & \textbf{110 } \\
Thunderbird & 522583  & 43546  & \textbf{18781 } & 261   & 283   & \textbf{107 } \\
Windows & 461847  & 14459  & \textbf{5721 } & 231   & 295   & \textbf{114 } \\
Linux & 444417  & 28972  & \textbf{9161 } & 222   & 287   & \textbf{82 } \\
Android & 430660  & 32574  & \textbf{17967 } & 215   & 256   & \textbf{115 } \\
HealthApp & 351681  & 16393  & \textbf{6980 } & 176   & 264   & \textbf{94 } \\
Apache & 364608  & 1549  & \textbf{950 } & 182   & 258   & \textbf{158 } \\
Proxifier & 494994	& 6724	& \textbf{4160} &	247 &	480 & \textbf{347} \\
OpenSSH & 490727  & 7921  & \textbf{4206 } & 245   & 317   & \textbf{162 } \\
OpenStack & 686037  & 14924  & \textbf{13798 } & 343   & 364   & \textbf{337 } \\
Mac   & 581290  & 109260  & \textbf{51002 } & 291   & 339   & \textbf{155} \\
\hline
Average & 469109  & 24020  & \textbf{11701} & 235   & 309   & \textbf{173} \\

\bottomrule
\end{tabular}
\end{table}
 
\subsubsection{Efficiency} In recent work, LLM-based parsers have concentrated on selecting suitable demonstrations for the query~\cite{xu2024divlog, jiang2023lilac}, which has led to demonstrations being significantly longer than the query itself.
Conversely, our approach improve the efficiency of using LLMs by adding more diverse logs to the query through \textit{batch-prompting}.
To compare the efficiency between \tool and other LLM-based parsers, we calculate the token consumption for all LLM-based log parsing baselines (i.e., DivLog and LILAC). The results are illustrated in Table~\ref{tab:efficiency}.
It is obvious that, in terms of total token consumption and average token consumption per invocation, our method achieves the best results on most datasets. For example, \tool exhibits the lowest \metricone and \metrictwo on 14 out of 16 datasets, with an average of 173 tokens per invocation, which is 26.4\% and 44.0\% lower than DivLog and LILAC, respectively. We notice that the total token consumption of DivLog is extremely high although its average token consumption per invocation is lower than LILAC. This is because DivLog queries LLMs for each log message, making parsing costly in practice. In contrast, \tool and LILAC adopt a caching mechanism to reduce the number of queries to LLMs, which significantly reduces the total token consumption. Moreover, \tool employs a batching -- querying mechanism to provide LLMs with more log messages per invocation, which further reduces the token consumption.


Overall, \tool does not require heuristic rules, handcrafted features, training process, or labeled data. Instead, \tool leverages latent commonalities and variabilities of log data to provide LLMs with better prompt context. Therefore, \tool can be directly applied to log parsing efficiently and effectively.


\begin{table}[h]
\centering
\caption{Ablation study results}
\vspace{-6pt}
\label{tab:ablation-study}
\resizebox{.93\linewidth}{!}{
\setlength{\tabcolsep}{1.5pt}
\renewcommand{\arraystretch}{1.1}
\begin{tabular}{lccc} 
\toprule
& \textbf{GA} & \textbf{MLA} & \textbf{ED}  \\
\hline
Full \tool  & 0.972   & 0.895  & 0.969  \\
w/o\textsubscript{partitioning} & 0.790\textsubscript{($\downarrow$23.0\%)} & 0.770\textsubscript{($\downarrow$16.2\%)} & 0.896\textsubscript{($\downarrow$8.1\%)} \\
w/o\textsubscript{caching} & 0.830\textsubscript{($\downarrow$14.1\%)} & 0.803\textsubscript{($\downarrow$11.5\%)} & 0.935\textsubscript{($\downarrow$3.6\%)} \\
w/o\textsubscript{batching} & 0.928\textsubscript{($\downarrow$4.7\%)} & 0.724\textsubscript{($\downarrow$23.6\%)} & 0.910\textsubscript{($\downarrow$6.5\%)} \\
\bottomrule
\end{tabular}
}
\end{table}

\subsection{\textbf{RQ2:} How do different modules contribute to \tool?}
This RQ gives a comprehensive explanation of each module's contribution. Table~\ref{tab:ablation-study} shows the results. It is clear that removing any of the three modules will affect performance to some extent.



\subsubsection{Partitioning} Intuitively, partitioning the logs is beneficial for the group accuracy, and this is indeed the case. Partitioning is the component that most significantly impacts grouping accuracy among the three components. Without it, the model's GA drops by 23.0\%.
Additionally, it also affects the message-level accuracy because the partitioning phase allows us to provide LLMs with commonalities within the input log data and correlate them with the log parsing task description. Without partitioning, the LLM input contains less commonalities, resulting in MLA decreased by 16.2\%.

\subsubsection{Caching} Due to the inherent limitations of the clustering method, logs belonging to the same template can be divided into different partitions. Within these partitions, our method achieves higher parsing accuracy for larger ones. When the caching module is removed, the results from these larger partitions cannot be used to guide the parsing of smaller ones. In other words, the smaller partitions are parsed independently, leading to poorer overall model performance. For example, removing caching decreases the GA and MLA of \tool by 14.1\% and 11.5\%, respectively, confirming the usefulness of the caching mechanism. 



\subsubsection{Batching} The proposed batching module is designed primarily to provide the LLM with diverse logs, expecting that the LLM can learn from the variability existing among the log data. The results demonstrate the importance of batching for the entire parsing process. Without batching, the MLA achieved by \tool significantly drops by 23.6\%, indicating that the LLM is less effective when the data provided has no diversity. This indicates that the batching module is the most crucial for \tool to achieve high parsing accuracy in terms of exact matching.

In summary, the ablation study confirms the usefulness of the each component in LogBatcher.

\begin{figure}[h]
  \centering
  \includegraphics[width=\linewidth]{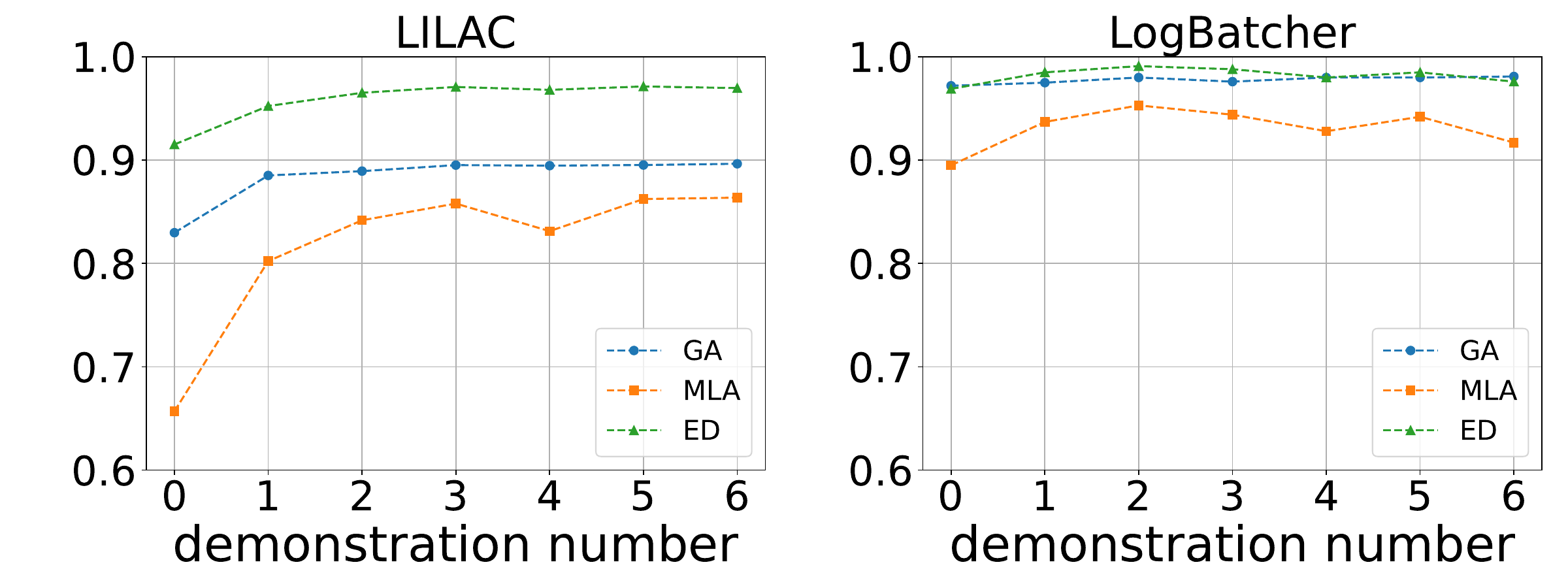 } 
  \caption{Average accuracy over different numbers of demonstrations}
  \label{fig:supervised_parser}
  \vspace{-12pt}
\end{figure}

\subsection{\textbf{RQ3:} How does \tool perform with with demonstrations?}

This RQ is to evaluate how \tool performs with demonstrations. 
As shown in Figure~\ref{fig:supervised_parser}, the demonstrations obtained through the sampling method adopted by~\cite{jiang2023lilac} indeed help improve performance. For example, the results of \tool in terms of MLA increase by 5.8\% when using only two demonstrations.
We notice that the performance varies to some extent when the number of demonstrations increases. In contrast, the performance of LogBatcher remains stable and high across different numbers of demonstrations. This is because LogBatcher already achieves good performance in the 0-shot setting, thus not requiring excessive reliance on examples. Furthermore, with the same number of examples, LogBatcher still significantly outperforms LILAC.


\subsection{\textbf{RQ4.} How do different settings affect \tool?}

In this RQ, we explore how different settings affect \tool from the following three aspects.

\subsubsection{Sampling Method} Selecting a sampling method involves determining which logs are grouped together into a batch. We examine three widely adopted sampling methods: similarity-based sampling, diversity-based sampling, and random sampling. To obtain a batch of similar logs, we use the approach from~\cite{cheng2023batch}, employing k-means clustering to identify and batch the most similar logs. For grouping diverse logs, we apply the Determinantal Point Process (DPP, a probabilistic model that favors diverse subsets by giving higher probabilities to dissimilar items)~\cite{kulesza2012dpp} method to ensure diversity. 
For random grouping, we sample logs randomly from the partition. Before sampling, we ensure that duplicates are removed from the log partition.
The result is shown in Table~\ref{tab:sampling_method}. 

\begin{table}[h]
  \centering
  \caption{Result with different sampling methods}
  \vspace{-6pt}
  \label{tab:sampling_method}
  \resizebox{.93\linewidth}{!}{
  \setlength{\tabcolsep}{1.5pt}
  \renewcommand{\arraystretch}{1.1}
  \begin{tabular}{lccc} 
  \toprule
  & \textbf{GA} & \textbf{MLA} & \textbf{ED}  \\
  \hline
  \tool  & 0.972   & 0.895  & 0.969   \\
  w/ \textsubscript{random sampling} & 0.963\textsubscript{($\downarrow$0.9\%)} & 0.890\textsubscript{($\downarrow$0.6\%)} & 0.964\textsubscript{($\downarrow$0.5\%)} \\
  w/ \textsubscript{similarity sampling} & 0.937\textsubscript{($\downarrow$3.7\%)} & 0.831\textsubscript{($\downarrow$7.7\%)} & 0.953\textsubscript{($\downarrow$1.7\%)} \\
  \bottomrule
  \end{tabular}
  }
  \end{table}

The diversity-based DPP algorithm achieves the best results because it provides LLM with sufficiently diverse logs. Random sampling only resulted in a slight decrease in performance because it can also select diverse logs to some extent. In contrast, similarity-based sampling decreases PA and MLA by 3.7\% and 7.7\% respectively. The results demonstrate that the diversity-based sampling method used in \tool is effective.  The results demonstrate that the diversity-based sampling method used in \tool is effective. 




\begin{table*}[t]
  \centering
  \caption{Comparison with LILAC on large-scale datasets from Loghub-2.0}
  \vspace{-6pt}
  \small
  \label{tab:big-scale}
  \setlength{\tabcolsep}{6pt}
  \renewcommand{\arraystretch}{1.1}
  \begin{tabular}{c|c|ccccc|ccccc} 
  \toprule
  \multicolumn{2}{c|}{} &  \multicolumn{5}{c|}{\textbf{LogBatcher}}    & \multicolumn{5}{c}{\textbf{LILAC}} \\
  \cline{1-12}
  Dataset & \#Log entries & GA    & MLA   & ED    & \metricone & \metrictwo & GA    & MLA   & ED    & \metricone & \metrictwo \\
  \cline{1-12}
  HDFS  & 11,167,740 & \textbf{1.000 } & \textbf{1.000} & \textbf{1.000} & \textbf{11646} & \textbf{233} & \textbf{1.000} & 0.999  & \textbf{1.000} & 16128  & 343  \\
  BGL   & 4,631,261 & \textbf{0.952} & 0.921  & 0.977  & \textbf{47428} & \textbf{121} & 0.910  & \textbf{0.975} & \textbf{0.996} & 86968  & 260  \\
  OpenStack & 207,632 & 0.990  & \textbf{0.979 } & \textbf{0.991 } & \textbf{15050} & \textbf{327} & \textbf{1.000}  & 0.970  & 0.988  & 17915  & 373  \\
  Zookeeper & 74,273 & 0.969  & \textbf{0.963 } & \textbf{0.993} & \textbf{10132} & \textbf{115} & \textbf{1.000} & 0.685  & 0.935  & 22787  & 253  \\
  \bottomrule
  \end{tabular}
  \end{table*}

\subsubsection{Batch Size} 
\begin{figure}[h]
  \centering
  \vspace{-6pt}
  \includegraphics[width=0.65\linewidth]{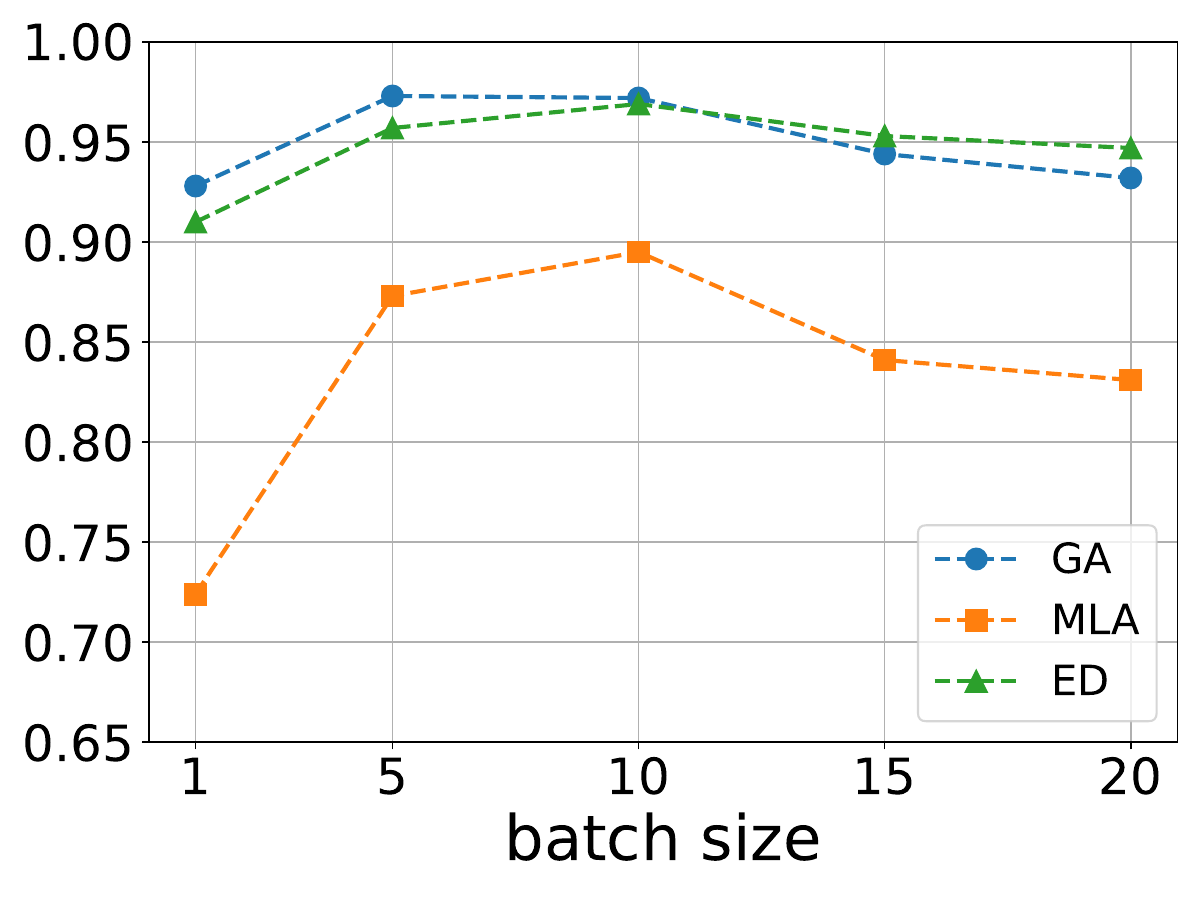}
  \vspace{-6pt}
  \caption{Average accuracy with different batch sizes}
  \vspace{-6pt}
  \label{fig:batchsize}
\end{figure}

To evaluate the impact of batch size on the performance of \tool, 
we select the batch size of 1, 5, 10, 15, and 20 (setting the batch size to 1 means removing the batching component). 
The results are shown in Figure~\ref{fig:batchsize}. It can be seen that when the batch size approaches 1, the performance drops significantly. The optimal batch size is found to be between 5 and 10. When the batch size exceeds 10, the performance of \tool slightly decreases. 
Considering larger batch sizes can lead to higher LLM invocation overhead, in our experiments we set the default batch size to 10. 

\subsubsection{LLM Selection}
In our experiments, we use ChatGPT (i.e., \textit{GPT-3.5-Turbo}) as the default LLM. 
We also evaluate the performance of our approach with different LLMs. 
We select two LLMs with different model parameter size, including Codellama 7B and Llama3 70B. Table~\ref{tab:model-backbone} shows the average metrics of \tool with different LLMs.
It is evident that \tool performs well even on a smaller LLM with a parameter count of 7B, achieving an average GA of 0.936 and MLA of 0.8. Overall, the larger the model's parameters, the better the performance. These findings demonstrate that LogBatcher can be effectively applied to various LLMs with robust performance. 


\begin{table}[h]
\centering
\caption{The performance of \tool when adopting different LLMs}
\vspace{-6pt}
\small
\label{tab:model-backbone}
\setlength{\tabcolsep}{6pt}
\renewcommand{\arraystretch}{1.1}
\begin{tabular}{lccc} 
\toprule
 & \textbf{GA} & \textbf{MLA} & \textbf{ED}  \\
\hline
CodeLlama (7B)  & 0.936   & 0.800  & 0.939  \\
Llama3 (70B)  & 0.925   & 0.853  & 0.966  \\
GPT-3.5-Turbo   & 0.972   & 0.895  & 0.969  \\
\bottomrule
\end{tabular}
 \vspace{-6pt}
\end{table}

\section{Discussion}
\label{sec:discussion}

\subsection{Practicality of \tool}

LogBatcher is designed for more practical log parsing with Large Language Models (LLMs).
Compared to other LLM-based log parsers, \tool does not need any training/fine-tuning process and labeling effort. Nevertheless, our experiments show that \tool can still achieve superior accuracy. Additionally, our method eliminates the need to select demonstrations for each query, significantly reducing the LLM invocation overhead. This leads to a notable improvement in cost-effectiveness and model robustness. 
It is also worth noting that \tool is compatible with many LLMs such as CodeLlama. 

A large amount of logs could be generated in production, so it is crucial to ensure that the log parser can perform online parsing, which means it can handle streaming log data. \tool can buffer a batch of streaming logs for parsing instead of processing each log individually. After the clustering and sorting process, the logs within the clusters will be parsed through the caching or querying stage, so no training process is needed. 

We also evaluate the applicability of \tool to large-scale datasets. 
Specifically, we selected four large-scale log datasets from Loghub-2.0~\cite{jiang2024large}, collected from real-world scenarios, with the number of data entries ranging from 74,000 to 11 million. We compared \tool with LILAC on this large-scale dataset. The results in Table~\ref{tab:big-scale} indicate that our method generally incurs less overhead, achieves higher efficiency, and performs better overall.


\subsection{Threats to Validity}
We have identified the following major threats to validity.

\textbf{Data Leakage.} The data leakage problem of LLM-based log parsers mainly manifests in two aspects: data leakage during the training process of the LLM itself, and the demonstrations during in-context learning disclosing the ground truth templates. 
Recent studies imply that there is a low probability of direct memorization of LLMs for the log parsing task as without in-context learning, the LLMs' performance significantly drops~\cite{jiang2023lilac,xu2024divlog}.  
Additionally, \tool does not require any training process or labeled data, no template is included in the prompt context, thus substantially eliminating the threat of leaking ground-truth templates within the prompt context. Overall, the probability of data leakage in our experiments is negligible.

\textbf{The Quality of Ground Truth Data.}
To fairly evaluate the effectiveness of \tool, the annotation quality for ground truth templates is critical. 
The datasets used in our experiments are from LogPai~\cite{jiang2024large}, which were manually lablelled by Zhu et al.~\cite{zhu2019tools}. 
It has been observed that the original labels have some errors and inconsistent labeling styles~\cite{khan2022guidelines}. To mitigate this threat, we use the datasets corrected by~\cite{khan2022guidelines}. Furthermore, we also perform experiments on another set of datasets Loghub-2.0~\cite{zhu2023loghub}. The results confirm that \tool is effective on both sets of datasets.

\section{Conclusion}
\label{sec:conclusion}
Log parsing is an important initial step for automated analysis of logs of large-scale software systems. To overcome the limitations of existing log parsers, we propose \method, a cost-effective LLM-based log parser that requires no training process or labeled demonstrations. \tool leverages latent characteristics of log data and reduces the LLM inference overhead 
by batching a group of logs. 
We have conducted extensive experiments on the public log dataset and the results show that \tool is effective and efficient for log parsing. We believe this {demonstration-free}, {training-free}, and {cost-effective} log parser has potential to make LLM-based log parsing more practical.


\noindent \textbf{Data Availability:} 
Our source code and experimental data are publicly available at
\url{https://anonymous.4open.science/r/LogBatcher}.


\balance
\bibliographystyle{ACM-Reference-Format}
\bibliography{sample-base}


\begin{thebibliography}{56}


\ifx \showCODEN    \undefined \def \showCODEN     #1{\unskip}     \fi
\ifx \showDOI      \undefined \def \showDOI       #1{#1}\fi
\ifx \showISBNx    \undefined \def \showISBNx     #1{\unskip}     \fi
\ifx \showISBNxiii \undefined \def \showISBNxiii  #1{\unskip}     \fi
\ifx \showISSN     \undefined \def \showISSN      #1{\unskip}     \fi
\ifx \showLCCN     \undefined \def \showLCCN      #1{\unskip}     \fi
\ifx \shownote     \undefined \def \shownote      #1{#1}          \fi
\ifx \showarticletitle \undefined \def \showarticletitle #1{#1}   \fi
\ifx \showURL      \undefined \def \showURL       {\relax}        \fi
\providecommand\bibfield[2]{#2}
\providecommand\bibinfo[2]{#2}
\providecommand\natexlab[1]{#1}
\providecommand\showeprint[2][]{arXiv:#2}

\bibitem[log(2023)]%
        {loghub2021}
 \bibinfo{year}{2023}\natexlab{}.
\newblock \bibinfo{title}{A large collection of system log datasets for AI-powered log analytics}.
\newblock
\newblock
\urldef\tempurl%
\url{https://github.com/logpai/loghub}
\showURL{%
Retrieved August 31, 2023 from \tempurl}


\bibitem[gpt(2024)]%
        {gpt-3.5-turbo}
 \bibinfo{year}{2024}\natexlab{}.
\newblock \bibinfo{title}{OpenAI ChatGPT}.
\newblock
\newblock
\urldef\tempurl%
\url{https://platform.openai.com/docs/models/gpt-3-5-turbo}
\showURL{%
Retrieved May 30, 2024 from \tempurl}


\bibitem[Cheng et~al\mbox{.}(2023)]%
        {cheng2023batch}
\bibfield{author}{\bibinfo{person}{Zhoujun Cheng}, \bibinfo{person}{Jungo Kasai}, {and} \bibinfo{person}{Tao Yu}.} \bibinfo{year}{2023}\natexlab{}.
\newblock \showarticletitle{Batch Prompting: Efficient Inference with Large Language Model APIs}. In \bibinfo{booktitle}{\emph{Proceedings of the 2023 Conference on Empirical Methods in Natural Language Processing: Industry Track}}. \bibinfo{pages}{792--810}.
\newblock


\bibitem[Dai et~al\mbox{.}(2020)]%
        {dai2020logram}
\bibfield{author}{\bibinfo{person}{Hetong Dai}, \bibinfo{person}{Heng Li}, \bibinfo{person}{Che~Shao Chen}, \bibinfo{person}{Weiyi Shang}, {and} \bibinfo{person}{Tse-Hsun Chen}.} \bibinfo{year}{2020}\natexlab{}.
\newblock \showarticletitle{Logram: Efficient log parsing using n-gram dictionaries}.
\newblock \bibinfo{journal}{\emph{IEEE Transactions on Software Engineering}} (\bibinfo{year}{2020}).
\newblock


\bibitem[Devlin et~al\mbox{.}(2019)]%
        {devlin2018bert}
\bibfield{author}{\bibinfo{person}{Jacob Devlin}, \bibinfo{person}{Ming-Wei Chang}, \bibinfo{person}{Kenton Lee}, {and} \bibinfo{person}{Kristina Toutanova}.} \bibinfo{year}{2019}\natexlab{}.
\newblock \showarticletitle{BERT: Pre-training of Deep Bidirectional Transformers for Language Understanding}. In \bibinfo{booktitle}{\emph{Proceedings of the 2019 Conference of the North American Chapter of the Association for Computational Linguistics: Human Language Technologies, Volume 1 (Long and Short Papers)}}. \bibinfo{pages}{4171--4186}.
\newblock


\bibitem[Du and Li(2016)]%
        {du2016spell}
\bibfield{author}{\bibinfo{person}{Min Du} {and} \bibinfo{person}{Feifei Li}.} \bibinfo{year}{2016}\natexlab{}.
\newblock \showarticletitle{Spell: Streaming parsing of system event logs}. In \bibinfo{booktitle}{\emph{2016 IEEE 16th International Conference on Data Mining (ICDM)}}. IEEE, \bibinfo{pages}{859--864}.
\newblock


\bibitem[Ester et~al\mbox{.}(1996)]%
        {ester1996density}
\bibfield{author}{\bibinfo{person}{Martin Ester}, \bibinfo{person}{Hans-Peter Kriegel}, \bibinfo{person}{J{\"o}rg Sander}, \bibinfo{person}{Xiaowei Xu}, {et~al\mbox{.}}} \bibinfo{year}{1996}\natexlab{}.
\newblock \showarticletitle{A density-based algorithm for discovering clusters in large spatial databases with noise}. In \bibinfo{booktitle}{\emph{kdd}}, Vol.~\bibinfo{volume}{96}. \bibinfo{pages}{226--231}.
\newblock


\bibitem[Fu et~al\mbox{.}(2009)]%
        {fu2009execution}
\bibfield{author}{\bibinfo{person}{Qiang Fu}, \bibinfo{person}{Jian-Guang Lou}, \bibinfo{person}{Yi Wang}, {and} \bibinfo{person}{Jiang Li}.} \bibinfo{year}{2009}\natexlab{}.
\newblock \showarticletitle{Execution anomaly detection in distributed systems through unstructured log analysis}. In \bibinfo{booktitle}{\emph{2009 ninth IEEE international conference on data mining}}. IEEE, \bibinfo{pages}{149--158}.
\newblock


\bibitem[Guo et~al\mbox{.}(2023)]%
        {guo2023images}
\bibfield{author}{\bibinfo{person}{Jiaxian Guo}, \bibinfo{person}{Junnan Li}, \bibinfo{person}{Dongxu Li}, \bibinfo{person}{Anthony Meng~Huat Tiong}, \bibinfo{person}{Boyang Li}, \bibinfo{person}{Dacheng Tao}, {and} \bibinfo{person}{Steven Hoi}.} \bibinfo{year}{2023}\natexlab{}.
\newblock \showarticletitle{From images to textual prompts: Zero-shot visual question answering with frozen large language models}. In \bibinfo{booktitle}{\emph{Proceedings of the IEEE/CVF Conference on Computer Vision and Pattern Recognition}}. \bibinfo{pages}{10867--10877}.
\newblock


\bibitem[He et~al\mbox{.}(2017)]%
        {he2017drain}
\bibfield{author}{\bibinfo{person}{Pinjia He}, \bibinfo{person}{Jieming Zhu}, \bibinfo{person}{Zibin Zheng}, {and} \bibinfo{person}{Michael~R Lyu}.} \bibinfo{year}{2017}\natexlab{}.
\newblock \showarticletitle{Drain: An online log parsing approach with fixed depth tree}. In \bibinfo{booktitle}{\emph{2017 IEEE International Conference on Web Services (ICWS)}}. IEEE, \bibinfo{pages}{33--40}.
\newblock


\bibitem[He et~al\mbox{.}(2021)]%
        {he2020survey}
\bibfield{author}{\bibinfo{person}{Shilin He}, \bibinfo{person}{Pinjia He}, \bibinfo{person}{Zhuangbin Chen}, \bibinfo{person}{Tianyi Yang}, \bibinfo{person}{Yuxin Su}, {and} \bibinfo{person}{Michael~R Lyu}.} \bibinfo{year}{2021}\natexlab{}.
\newblock \showarticletitle{A survey on automated log analysis for reliability engineering}.
\newblock \bibinfo{journal}{\emph{ACM computing surveys (CSUR)}} \bibinfo{volume}{54}, \bibinfo{number}{6} (\bibinfo{year}{2021}), \bibinfo{pages}{1--37}.
\newblock


\bibitem[Hidv{\'e}gi et~al\mbox{.}(2024)]%
        {hidvegi2024cigar}
\bibfield{author}{\bibinfo{person}{D{\'a}vid Hidv{\'e}gi}, \bibinfo{person}{Khashayar Etemadi}, \bibinfo{person}{Sofia Bobadilla}, {and} \bibinfo{person}{Martin Monperrus}.} \bibinfo{year}{2024}\natexlab{}.
\newblock \showarticletitle{CigaR: Cost-efficient Program Repair with LLMs}.
\newblock \bibinfo{journal}{\emph{arXiv preprint arXiv:2402.06598}} (\bibinfo{year}{2024}).
\newblock


\bibitem[Jia et~al\mbox{.}(2017)]%
        {jia2017logsed}
\bibfield{author}{\bibinfo{person}{Tong Jia}, \bibinfo{person}{Lin Yang}, \bibinfo{person}{Pengfei Chen}, \bibinfo{person}{Ying Li}, \bibinfo{person}{Fanjing Meng}, {and} \bibinfo{person}{Jingmin Xu}.} \bibinfo{year}{2017}\natexlab{}.
\newblock \showarticletitle{Logsed: Anomaly diagnosis through mining time-weighted control flow graph in logs}. In \bibinfo{booktitle}{\emph{2017 IEEE 10th International Conference on Cloud Computing (CLOUD)}}. IEEE, \bibinfo{pages}{447--455}.
\newblock


\bibitem[Jiang et~al\mbox{.}(2024a)]%
        {jiang2023lilac}
\bibfield{author}{\bibinfo{person}{Zhihan Jiang}, \bibinfo{person}{Jinyang Liu}, \bibinfo{person}{Zhuangbin Chen}, \bibinfo{person}{Yichen Li}, \bibinfo{person}{Junjie Huang}, \bibinfo{person}{Yintong Huo}, \bibinfo{person}{Pinjia He}, \bibinfo{person}{Jiazhen Gu}, {and} \bibinfo{person}{Michael~R. Lyu}.} \bibinfo{year}{2024}\natexlab{a}.
\newblock \bibinfo{title}{LILAC: Log Parsing using LLMs with Adaptive Parsing Cache}.
\newblock
\newblock
\showeprint[arxiv]{2310.01796}~[cs.SE]


\bibitem[Jiang et~al\mbox{.}(2024b)]%
        {jiang2024large}
\bibfield{author}{\bibinfo{person}{Zhihan Jiang}, \bibinfo{person}{Jinyang Liu}, \bibinfo{person}{Junjie Huang}, \bibinfo{person}{Yichen Li}, \bibinfo{person}{Yintong Huo}, \bibinfo{person}{Jiazhen Gu}, \bibinfo{person}{Zhuangbin Chen}, \bibinfo{person}{Jieming Zhu}, {and} \bibinfo{person}{Michael~R Lyu}.} \bibinfo{year}{2024}\natexlab{b}.
\newblock \showarticletitle{A Large-Scale Evaluation for Log Parsing Techniques: How Far Are We?}. In \bibinfo{booktitle}{\emph{Proceedings of the 33rd ACM SIGSOFT International Symposium on Software Testing and Analysis}}.
\newblock


\bibitem[Jiang et~al\mbox{.}(2008)]%
        {jiang2008abstracting}
\bibfield{author}{\bibinfo{person}{Zhen~Ming Jiang}, \bibinfo{person}{Ahmed~E Hassan}, \bibinfo{person}{Parminder Flora}, {and} \bibinfo{person}{Gilbert Hamann}.} \bibinfo{year}{2008}\natexlab{}.
\newblock \showarticletitle{Abstracting execution logs to execution events for enterprise applications (short paper)}. In \bibinfo{booktitle}{\emph{2008 The Eighth International Conference on Quality Software}}. IEEE, \bibinfo{pages}{181--186}.
\newblock


\bibitem[Kabinna et~al\mbox{.}(2018)]%
        {kabinna2018examining}
\bibfield{author}{\bibinfo{person}{Suhas Kabinna}, \bibinfo{person}{Cor-Paul Bezemer}, \bibinfo{person}{Weiyi Shang}, \bibinfo{person}{Mark~D Syer}, {and} \bibinfo{person}{Ahmed~E Hassan}.} \bibinfo{year}{2018}\natexlab{}.
\newblock \showarticletitle{Examining the stability of logging statements}.
\newblock \bibinfo{journal}{\emph{Empirical Software Engineering}} \bibinfo{volume}{23}, \bibinfo{number}{1} (\bibinfo{year}{2018}), \bibinfo{pages}{290--333}.
\newblock


\bibitem[Khan et~al\mbox{.}(2022)]%
        {khan2022guidelines}
\bibfield{author}{\bibinfo{person}{Zanis~Ali Khan}, \bibinfo{person}{Donghwan Shin}, \bibinfo{person}{Domenico Bianculli}, {and} \bibinfo{person}{Lionel Briand}.} \bibinfo{year}{2022}\natexlab{}.
\newblock \showarticletitle{Guidelines for assessing the accuracy of log message template identification techniques}. In \bibinfo{booktitle}{\emph{Proceedings of the 44th International Conference on Software Engineering}}. \bibinfo{pages}{1095--1106}.
\newblock


\bibitem[Khot et~al\mbox{.}(2023)]%
        {khot2022decomposed}
\bibfield{author}{\bibinfo{person}{Tushar Khot}, \bibinfo{person}{Harsh Trivedi}, \bibinfo{person}{Matthew Finlayson}, \bibinfo{person}{Yao Fu}, \bibinfo{person}{Kyle Richardson}, \bibinfo{person}{Peter Clark}, {and} \bibinfo{person}{Ashish Sabharwal}.} \bibinfo{year}{2023}\natexlab{}.
\newblock \showarticletitle{Decomposed Prompting: A Modular Approach for Solving Complex Tasks}. In \bibinfo{booktitle}{\emph{The Eleventh International Conference on Learning Representations}}.
\newblock


\bibitem[Kulesza et~al\mbox{.}(2012)]%
        {kulesza2012dpp}
\bibfield{author}{\bibinfo{person}{Alex Kulesza}, \bibinfo{person}{Ben Taskar}, {et~al\mbox{.}}} \bibinfo{year}{2012}\natexlab{}.
\newblock \showarticletitle{Determinantal point processes for machine learning}.
\newblock \bibinfo{journal}{\emph{Foundations and Trends{\textregistered} in Machine Learning}} \bibinfo{volume}{5}, \bibinfo{number}{2--3} (\bibinfo{year}{2012}), \bibinfo{pages}{123--286}.
\newblock


\bibitem[Le and Zhang(2021)]%
        {le2021log}
\bibfield{author}{\bibinfo{person}{Van-Hoang Le} {and} \bibinfo{person}{Hongyu Zhang}.} \bibinfo{year}{2021}\natexlab{}.
\newblock \showarticletitle{Log-based Anomaly Detection Without Log Parsing}. In \bibinfo{booktitle}{\emph{2021 36th IEEE/ACM International Conference on Automated Software Engineering (ASE)}}. \bibinfo{pages}{492--504}.
\newblock


\bibitem[Le and Zhang(2022)]%
        {le2022log}
\bibfield{author}{\bibinfo{person}{Van-Hoang Le} {and} \bibinfo{person}{Hongyu Zhang}.} \bibinfo{year}{2022}\natexlab{}.
\newblock \showarticletitle{Log-based anomaly detection with deep learning: How far are we?}. In \bibinfo{booktitle}{\emph{Proceedings of the 44th International Conference on Software Engineering}}. \bibinfo{pages}{1356--1367}.
\newblock


\bibitem[Le and Zhang(2023a)]%
        {le2023evaluation}
\bibfield{author}{\bibinfo{person}{Van-Hoang Le} {and} \bibinfo{person}{Hongyu Zhang}.} \bibinfo{year}{2023}\natexlab{a}.
\newblock \showarticletitle{Log Parsing: How Far Can ChatGPT Go?}. In \bibinfo{booktitle}{\emph{2023 38th IEEE/ACM International Conference on Automated Software Engineering (ASE)}}. IEEE, \bibinfo{pages}{1699--1704}.
\newblock


\bibitem[Le and Zhang(2023b)]%
        {le2023log}
\bibfield{author}{\bibinfo{person}{Van-Hoang Le} {and} \bibinfo{person}{Hongyu Zhang}.} \bibinfo{year}{2023}\natexlab{b}.
\newblock \showarticletitle{Log parsing with prompt-based few-shot learning}. In \bibinfo{booktitle}{\emph{2023 IEEE/ACM 45th International Conference on Software Engineering (ICSE)}}. IEEE, \bibinfo{pages}{2438--2449}.
\newblock


\bibitem[Lewis et~al\mbox{.}(2020)]%
        {lewis2019bart}
\bibfield{author}{\bibinfo{person}{Mike Lewis}, \bibinfo{person}{Yinhan Liu}, \bibinfo{person}{Naman Goyal}, \bibinfo{person}{Marjan Ghazvininejad}, \bibinfo{person}{Abdelrahman Mohamed}, \bibinfo{person}{Omer Levy}, \bibinfo{person}{Veselin Stoyanov}, {and} \bibinfo{person}{Luke Zettlemoyer}.} \bibinfo{year}{2020}\natexlab{}.
\newblock \showarticletitle{BART: Denoising Sequence-to-Sequence Pre-training for Natural Language Generation, Translation, and Comprehension}. In \bibinfo{booktitle}{\emph{Proceedings of the 58th Annual Meeting of the Association for Computational Linguistics}}. \bibinfo{pages}{7871--7880}.
\newblock


\bibitem[Li et~al\mbox{.}(2024)]%
        {li2024logshrink}
\bibfield{author}{\bibinfo{person}{Xiaoyun Li}, \bibinfo{person}{Hongyu Zhang}, \bibinfo{person}{Van-Hoang Le}, {and} \bibinfo{person}{Pengfei Chen}.} \bibinfo{year}{2024}\natexlab{}.
\newblock \showarticletitle{Logshrink: Effective log compression by leveraging commonality and variability of log data}. In \bibinfo{booktitle}{\emph{Proceedings of the 46th IEEE/ACM International Conference on Software Engineering}}. \bibinfo{pages}{1--12}.
\newblock


\bibitem[Liu et~al\mbox{.}(2023)]%
        {liu2023pre}
\bibfield{author}{\bibinfo{person}{Pengfei Liu}, \bibinfo{person}{Weizhe Yuan}, \bibinfo{person}{Jinlan Fu}, \bibinfo{person}{Zhengbao Jiang}, \bibinfo{person}{Hiroaki Hayashi}, {and} \bibinfo{person}{Graham Neubig}.} \bibinfo{year}{2023}\natexlab{}.
\newblock \showarticletitle{Pre-train, prompt, and predict: A systematic survey of prompting methods in natural language processing}.
\newblock \bibinfo{journal}{\emph{Comput. Surveys}} \bibinfo{volume}{55}, \bibinfo{number}{9} (\bibinfo{year}{2023}), \bibinfo{pages}{1--35}.
\newblock


\bibitem[Liu et~al\mbox{.}(2022)]%
        {liu2022uniparser}
\bibfield{author}{\bibinfo{person}{Yudong Liu}, \bibinfo{person}{Xu Zhang}, \bibinfo{person}{Shilin He}, \bibinfo{person}{Hongyu Zhang}, \bibinfo{person}{Liqun Li}, \bibinfo{person}{Yu Kang}, \bibinfo{person}{Yong Xu}, \bibinfo{person}{Minghua Ma}, \bibinfo{person}{Qingwei Lin}, \bibinfo{person}{Yingnong Dang}, {et~al\mbox{.}}} \bibinfo{year}{2022}\natexlab{}.
\newblock \showarticletitle{Uniparser: A unified log parser for heterogeneous log data}. In \bibinfo{booktitle}{\emph{Proceedings of the ACM Web Conference 2022}}. \bibinfo{pages}{1893--1901}.
\newblock


\bibitem[Ma et~al\mbox{.}(2024)]%
        {Ma_2024}
\bibfield{author}{\bibinfo{person}{Zeyang Ma}, \bibinfo{person}{An~Ran Chen}, \bibinfo{person}{Dong~Jae Kim}, \bibinfo{person}{Tse-Hsun Chen}, {and} \bibinfo{person}{Shaowei Wang}.} \bibinfo{year}{2024}\natexlab{}.
\newblock \showarticletitle{LLMParser: An Exploratory Study on Using Large Language Models for Log Parsing}. In \bibinfo{booktitle}{\emph{Proceedings of the IEEE/ACM 46th International Conference on Software Engineering}} \emph{(\bibinfo{series}{ICSE ’24})}. \bibinfo{publisher}{ACM}.
\newblock
\urldef\tempurl%
\url{https://doi.org/10.1145/3597503.3639150}
\showDOI{\tempurl}


\bibitem[Makanju et~al\mbox{.}(2009)]%
        {makanju2009clustering}
\bibfield{author}{\bibinfo{person}{Adetokunbo~AO Makanju}, \bibinfo{person}{A~Nur Zincir-Heywood}, {and} \bibinfo{person}{Evangelos~E Milios}.} \bibinfo{year}{2009}\natexlab{}.
\newblock \showarticletitle{Clustering event logs using iterative partitioning}. In \bibinfo{booktitle}{\emph{Proceedings of the 15th ACM SIGKDD international conference on Knowledge discovery and data mining}}. \bibinfo{pages}{1255--1264}.
\newblock


\bibitem[Marzal and Vidal(1993)]%
        {marzal1993computation}
\bibfield{author}{\bibinfo{person}{Andres Marzal} {and} \bibinfo{person}{Enrique Vidal}.} \bibinfo{year}{1993}\natexlab{}.
\newblock \showarticletitle{Computation of normalized edit distance and applications}.
\newblock \bibinfo{journal}{\emph{IEEE transactions on pattern analysis and machine intelligence}} \bibinfo{volume}{15}, \bibinfo{number}{9} (\bibinfo{year}{1993}), \bibinfo{pages}{926--932}.
\newblock


\bibitem[Mi et~al\mbox{.}(2013)]%
        {mi2013toward}
\bibfield{author}{\bibinfo{person}{Haibo Mi}, \bibinfo{person}{Huaimin Wang}, \bibinfo{person}{Yangfan Zhou}, \bibinfo{person}{Michael Rung-Tsong Lyu}, {and} \bibinfo{person}{Hua Cai}.} \bibinfo{year}{2013}\natexlab{}.
\newblock \showarticletitle{Toward fine-grained, unsupervised, scalable performance diagnosis for production cloud computing systems}.
\newblock \bibinfo{journal}{\emph{IEEE Transactions on Parallel and Distributed Systems}} \bibinfo{volume}{24}, \bibinfo{number}{6} (\bibinfo{year}{2013}), \bibinfo{pages}{1245--1255}.
\newblock


\bibitem[Milajerdi et~al\mbox{.}(2019)]%
        {milajerdi2019poirot}
\bibfield{author}{\bibinfo{person}{Sadegh~M Milajerdi}, \bibinfo{person}{Birhanu Eshete}, \bibinfo{person}{Rigel Gjomemo}, {and} \bibinfo{person}{VN Venkatakrishnan}.} \bibinfo{year}{2019}\natexlab{}.
\newblock \showarticletitle{Poirot: Aligning attack behavior with kernel audit records for cyber threat hunting}. In \bibinfo{booktitle}{\emph{Proceedings of the 2019 ACM SIGSAC conference on computer and communications security}}. \bibinfo{pages}{1795--1812}.
\newblock


\bibitem[Nagappan and Vouk(2010)]%
        {nagappan2010abstracting}
\bibfield{author}{\bibinfo{person}{Meiyappan Nagappan} {and} \bibinfo{person}{Mladen~A Vouk}.} \bibinfo{year}{2010}\natexlab{}.
\newblock \showarticletitle{Abstracting log lines to log event types for mining software system logs}. In \bibinfo{booktitle}{\emph{2010 7th IEEE Working Conference on Mining Software Repositories (MSR 2010)}}. IEEE, \bibinfo{pages}{114--117}.
\newblock


\bibitem[Nagappan et~al\mbox{.}(2009)]%
        {nagappan2009efficiently}
\bibfield{author}{\bibinfo{person}{Meiyappan Nagappan}, \bibinfo{person}{Kesheng Wu}, {and} \bibinfo{person}{Mladen~A Vouk}.} \bibinfo{year}{2009}\natexlab{}.
\newblock \showarticletitle{Efficiently extracting operational profiles from execution logs using suffix arrays}. In \bibinfo{booktitle}{\emph{2009 20th International Symposium on Software Reliability Engineering}}. IEEE, \bibinfo{pages}{41--50}.
\newblock


\bibitem[Nedelkoski et~al\mbox{.}(2020)]%
        {nedelkoski2020self}
\bibfield{author}{\bibinfo{person}{Sasho Nedelkoski}, \bibinfo{person}{Jasmin Bogatinovski}, \bibinfo{person}{Alexander Acker}, \bibinfo{person}{Jorge Cardoso}, {and} \bibinfo{person}{Odej Kao}.} \bibinfo{year}{2020}\natexlab{}.
\newblock \showarticletitle{Self-attentive classification-based anomaly detection in unstructured logs}. In \bibinfo{booktitle}{\emph{2020 IEEE International Conference on Data Mining (ICDM)}}. IEEE, \bibinfo{pages}{1196--1201}.
\newblock


\bibitem[OpenAI(2023)]%
        {openai2023gpt4}
\bibfield{author}{\bibinfo{person}{OpenAI}.} \bibinfo{year}{2023}\natexlab{}.
\newblock \showarticletitle{GPT-4 Technical Report}.
\newblock \bibinfo{journal}{\emph{ArXiv}}  \bibinfo{volume}{abs/2303.08774} (\bibinfo{year}{2023}).
\newblock


\bibitem[Oprea et~al\mbox{.}(2015)]%
        {oprea2015detection}
\bibfield{author}{\bibinfo{person}{Alina Oprea}, \bibinfo{person}{Zhou Li}, \bibinfo{person}{Ting-Fang Yen}, \bibinfo{person}{Sang~H Chin}, {and} \bibinfo{person}{Sumayah Alrwais}.} \bibinfo{year}{2015}\natexlab{}.
\newblock \showarticletitle{Detection of early-stage enterprise infection by mining large-scale log data}. In \bibinfo{booktitle}{\emph{2015 45th Annual IEEE/IFIP International Conference on Dependable Systems and Networks}}. IEEE, \bibinfo{pages}{45--56}.
\newblock


\bibitem[Pedregosa et~al\mbox{.}(2011)]%
        {pedregosa2011scikit}
\bibfield{author}{\bibinfo{person}{Fabian Pedregosa}, \bibinfo{person}{Ga{\"e}l Varoquaux}, \bibinfo{person}{Alexandre Gramfort}, \bibinfo{person}{Vincent Michel}, \bibinfo{person}{Bertrand Thirion}, \bibinfo{person}{Olivier Grisel}, \bibinfo{person}{Mathieu Blondel}, \bibinfo{person}{Peter Prettenhofer}, \bibinfo{person}{Ron Weiss}, \bibinfo{person}{Vincent Dubourg}, {et~al\mbox{.}}} \bibinfo{year}{2011}\natexlab{}.
\newblock \showarticletitle{Scikit-learn: Machine learning in Python}.
\newblock \bibinfo{journal}{\emph{the Journal of machine Learning research}}  \bibinfo{volume}{12} (\bibinfo{year}{2011}), \bibinfo{pages}{2825--2830}.
\newblock


\bibitem[Petrescu et~al\mbox{.}(2023)]%
        {petrescu2023log}
\bibfield{author}{\bibinfo{person}{Stefan Petrescu}, \bibinfo{person}{Floris Den~Hengst}, \bibinfo{person}{Alexandru Uta}, {and} \bibinfo{person}{Jan~S Rellermeyer}.} \bibinfo{year}{2023}\natexlab{}.
\newblock \showarticletitle{Log parsing evaluation in the era of modern software systems}. In \bibinfo{booktitle}{\emph{2023 IEEE 34th International Symposium on Software Reliability Engineering (ISSRE)}}. IEEE, \bibinfo{pages}{379--390}.
\newblock


\bibitem[Salton and Buckley(1988)]%
        {salton1988term}
\bibfield{author}{\bibinfo{person}{Gerard Salton} {and} \bibinfo{person}{Christopher Buckley}.} \bibinfo{year}{1988}\natexlab{}.
\newblock \showarticletitle{Term-weighting approaches in automatic text retrieval}.
\newblock \bibinfo{journal}{\emph{Information processing \& management}} \bibinfo{volume}{24}, \bibinfo{number}{5} (\bibinfo{year}{1988}), \bibinfo{pages}{513--523}.
\newblock


\bibitem[Schubert et~al\mbox{.}(2017)]%
        {schubert2017dbscan}
\bibfield{author}{\bibinfo{person}{Erich Schubert}, \bibinfo{person}{J{\"o}rg Sander}, \bibinfo{person}{Martin Ester}, \bibinfo{person}{Hans~Peter Kriegel}, {and} \bibinfo{person}{Xiaowei Xu}.} \bibinfo{year}{2017}\natexlab{}.
\newblock \showarticletitle{DBSCAN revisited, revisited: why and how you should (still) use DBSCAN}.
\newblock \bibinfo{journal}{\emph{ACM Transactions on Database Systems (TODS)}} \bibinfo{volume}{42}, \bibinfo{number}{3} (\bibinfo{year}{2017}), \bibinfo{pages}{1--21}.
\newblock


\bibitem[Shao et~al\mbox{.}(2023)]%
        {shao2023prompting}
\bibfield{author}{\bibinfo{person}{Zhenwei Shao}, \bibinfo{person}{Zhou Yu}, \bibinfo{person}{Meng Wang}, {and} \bibinfo{person}{Jun Yu}.} \bibinfo{year}{2023}\natexlab{}.
\newblock \showarticletitle{Prompting large language models with answer heuristics for knowledge-based visual question answering}. In \bibinfo{booktitle}{\emph{Proceedings of the IEEE/CVF Conference on Computer Vision and Pattern Recognition}}. \bibinfo{pages}{14974--14983}.
\newblock


\bibitem[Shima(2016)]%
        {shima2016length}
\bibfield{author}{\bibinfo{person}{Keiichi Shima}.} \bibinfo{year}{2016}\natexlab{}.
\newblock \showarticletitle{Length matters: Clustering system log messages using length of words}.
\newblock \bibinfo{journal}{\emph{arXiv preprint arXiv:1611.03213}} (\bibinfo{year}{2016}).
\newblock


\bibitem[Tang et~al\mbox{.}(2011)]%
        {tang2011logsig}
\bibfield{author}{\bibinfo{person}{Liang Tang}, \bibinfo{person}{Tao Li}, {and} \bibinfo{person}{Chang-Shing Perng}.} \bibinfo{year}{2011}\natexlab{}.
\newblock \showarticletitle{LogSig: Generating system events from raw textual logs}. In \bibinfo{booktitle}{\emph{Proceedings of the 20th ACM international conference on Information and knowledge management}}. \bibinfo{pages}{785--794}.
\newblock


\bibitem[Vaarandi and Pihelgas(2015)]%
        {vaarandi2015logcluster}
\bibfield{author}{\bibinfo{person}{Risto Vaarandi} {and} \bibinfo{person}{Mauno Pihelgas}.} \bibinfo{year}{2015}\natexlab{}.
\newblock \showarticletitle{Logcluster-a data clustering and pattern mining algorithm for event logs}. In \bibinfo{booktitle}{\emph{2015 11th International conference on network and service management (CNSM)}}. IEEE, \bibinfo{pages}{1--7}.
\newblock


\bibitem[Wang et~al\mbox{.}(2022)]%
        {wang2022spine}
\bibfield{author}{\bibinfo{person}{Xuheng Wang}, \bibinfo{person}{Xu Zhang}, \bibinfo{person}{Liqun Li}, \bibinfo{person}{Shilin He}, \bibinfo{person}{Hongyu Zhang}, \bibinfo{person}{Yudong Liu}, \bibinfo{person}{Lingling Zheng}, \bibinfo{person}{Yu Kang}, \bibinfo{person}{Qingwei Lin}, \bibinfo{person}{Yingnong Dang}, {et~al\mbox{.}}} \bibinfo{year}{2022}\natexlab{}.
\newblock \showarticletitle{SPINE: a scalable log parser with feedback guidance}. In \bibinfo{booktitle}{\emph{Proceedings of the 30th ACM Joint European Software Engineering Conference and Symposium on the Foundations of Software Engineering}}. \bibinfo{pages}{1198--1208}.
\newblock


\bibitem[Wei et~al\mbox{.}(2022)]%
        {wei2022chain}
\bibfield{author}{\bibinfo{person}{Jason Wei}, \bibinfo{person}{Xuezhi Wang}, \bibinfo{person}{Dale Schuurmans}, \bibinfo{person}{Maarten Bosma}, \bibinfo{person}{Fei Xia}, \bibinfo{person}{Ed Chi}, \bibinfo{person}{Quoc~V Le}, \bibinfo{person}{Denny Zhou}, {et~al\mbox{.}}} \bibinfo{year}{2022}\natexlab{}.
\newblock \showarticletitle{Chain-of-thought prompting elicits reasoning in large language models}.
\newblock \bibinfo{journal}{\emph{Advances in neural information processing systems}}  \bibinfo{volume}{35} (\bibinfo{year}{2022}), \bibinfo{pages}{24824--24837}.
\newblock


\bibitem[Xia et~al\mbox{.}(2024)]%
        {xia2024fuzz4all}
\bibfield{author}{\bibinfo{person}{Chunqiu~Steven Xia}, \bibinfo{person}{Matteo Paltenghi}, \bibinfo{person}{Jia Le~Tian}, \bibinfo{person}{Michael Pradel}, {and} \bibinfo{person}{Lingming Zhang}.} \bibinfo{year}{2024}\natexlab{}.
\newblock \showarticletitle{Fuzz4all: Universal fuzzing with large language models}. In \bibinfo{booktitle}{\emph{Proceedings of the IEEE/ACM 46th International Conference on Software Engineering}}. \bibinfo{pages}{1--13}.
\newblock


\bibitem[Xu et~al\mbox{.}(2024)]%
        {xu2024divlog}
\bibfield{author}{\bibinfo{person}{Junjielong Xu}, \bibinfo{person}{Ruichun Yang}, \bibinfo{person}{Yintong Huo}, \bibinfo{person}{Chengyu Zhang}, {and} \bibinfo{person}{Pinjia He}.} \bibinfo{year}{2024}\natexlab{}.
\newblock \showarticletitle{DivLog: Log Parsing with Prompt Enhanced In-Context Learning}. In \bibinfo{booktitle}{\emph{Proceedings of the IEEE/ACM 46th International Conference on Software Engineering}}. \bibinfo{pages}{1--12}.
\newblock


\bibitem[Xu et~al\mbox{.}(2009)]%
        {xu2009detecting}
\bibfield{author}{\bibinfo{person}{Wei Xu}, \bibinfo{person}{Ling Huang}, \bibinfo{person}{Armando Fox}, \bibinfo{person}{David Patterson}, {and} \bibinfo{person}{Michael~I Jordan}.} \bibinfo{year}{2009}\natexlab{}.
\newblock \showarticletitle{Detecting large-scale system problems by mining console logs}. In \bibinfo{booktitle}{\emph{Proceedings of the ACM SIGOPS 22nd symposium on Operating systems principles}}. \bibinfo{pages}{117--132}.
\newblock


\bibitem[Yu et~al\mbox{.}(2023)]%
        {yu2023brain}
\bibfield{author}{\bibinfo{person}{Siyu Yu}, \bibinfo{person}{Pinjia He}, \bibinfo{person}{Ningjiang Chen}, {and} \bibinfo{person}{Yifan Wu}.} \bibinfo{year}{2023}\natexlab{}.
\newblock \showarticletitle{Brain: Log parsing with bidirectional parallel tree}.
\newblock \bibinfo{journal}{\emph{IEEE Transactions on Services Computing}} (\bibinfo{year}{2023}).
\newblock


\bibitem[Zhang et~al\mbox{.}(2024)]%
        {zhang2024lemur}
\bibfield{author}{\bibinfo{person}{Wei Zhang}, \bibinfo{person}{Hongcheng Guo}, \bibinfo{person}{Anjie Le}, \bibinfo{person}{Jian Yang}, \bibinfo{person}{Jiaheng Liu}, \bibinfo{person}{Zhoujun Li}, \bibinfo{person}{Tieqiao Zheng}, \bibinfo{person}{Shi Xu}, \bibinfo{person}{Runqiang Zang}, \bibinfo{person}{Liangfan Zheng}, {and} \bibinfo{person}{Bo Zhang}.} \bibinfo{year}{2024}\natexlab{}.
\newblock \bibinfo{title}{Lemur: Log Parsing with Entropy Sampling and Chain-of-Thought Merging}.
\newblock
\newblock
\showeprint[arxiv]{2402.18205}~[cs.SE]


\bibitem[Zhang et~al\mbox{.}(2019)]%
        {zhang2019robust}
\bibfield{author}{\bibinfo{person}{Xu Zhang}, \bibinfo{person}{Yong Xu}, \bibinfo{person}{Qingwei Lin}, \bibinfo{person}{Bo Qiao}, \bibinfo{person}{Hongyu Zhang}, \bibinfo{person}{Yingnong Dang}, \bibinfo{person}{Chunyu Xie}, \bibinfo{person}{Xinsheng Yang}, \bibinfo{person}{Qian Cheng}, \bibinfo{person}{Ze Li}, {et~al\mbox{.}}} \bibinfo{year}{2019}\natexlab{}.
\newblock \showarticletitle{Robust log-based anomaly detection on unstable log data}. In \bibinfo{booktitle}{\emph{Proceedings of the 2019 27th ACM Joint Meeting on European Software Engineering Conference and Symposium on the Foundations of Software Engineering}}. \bibinfo{pages}{807--817}.
\newblock


\bibitem[Zhu et~al\mbox{.}(2023)]%
        {zhu2023loghub}
\bibfield{author}{\bibinfo{person}{Jieming Zhu}, \bibinfo{person}{Shilin He}, \bibinfo{person}{Pinjia He}, \bibinfo{person}{Jinyang Liu}, {and} \bibinfo{person}{Michael~R. Lyu}.} \bibinfo{year}{2023}\natexlab{}.
\newblock \bibinfo{title}{Loghub: A Large Collection of System Log Datasets for AI-driven Log Analytics}.
\newblock
\newblock
\showeprint[arxiv]{2008.06448}~[cs.SE]


\bibitem[Zhu et~al\mbox{.}(2019)]%
        {zhu2019tools}
\bibfield{author}{\bibinfo{person}{Jieming Zhu}, \bibinfo{person}{Shilin He}, \bibinfo{person}{Jinyang Liu}, \bibinfo{person}{Pinjia He}, \bibinfo{person}{Qi Xie}, \bibinfo{person}{Zibin Zheng}, {and} \bibinfo{person}{Michael~R Lyu}.} \bibinfo{year}{2019}\natexlab{}.
\newblock \showarticletitle{Tools and benchmarks for automated log parsing}. In \bibinfo{booktitle}{\emph{2019 IEEE/ACM 41st International Conference on Software Engineering: Software Engineering in Practice (ICSE-SEIP)}}. IEEE, \bibinfo{pages}{121--130}.
\newblock


\end{thebibliography}
\balance

\end{document}